\documentclass[%
superscriptaddress,
preprint,
 amsmath,amssymb,
 aps,
floatfix %
]{revtex4-2}

\usepackage{graphicx,color}% Include figure files %★arXiv, PRCに投稿する際にはこちらを使用する！★
\usepackage{dcolumn}% Align table columns on decimal point
\usepackage{bm}% bold math
\usepackage{multirow}

\begin{document}
%\preprint{APS/123-QED}
\title{Two unitary limits in low-energy $s$-wave neutron scattering on superfluid nuclei}% Force line breaks with \\
%\thanks{A footnote to the article title}%

\author{Yoshihiko Kobayashi}
%\altaffiliation[Also at ]{Physics Department, XYZ University.}%Lines break automatically or can be forced with \\
%\author{Second Author}%
\email{yoshikoba@oita-u.ac.jp}
\affiliation{Faculty of Education, Oita University, Oita 870-1192, Japan}%

%\collaboration{MUSO Collaboration}%\noaffiliation

\author{Masayuki Matsuo}
%\homepage{http://www.Second.institution.edu/~Charlie.Author}
\email{matsuo@phys.sc.niigata-u.ac.jp}
\affiliation{Faculty of Science, Niigata University, Niigata 950-2181, Japan}%
\affiliation{Research Center for Nuclear Physics, Osaka University, Ibaraki 567-0047, Japan}
%\affiliation{
% Third institution, the second for Charlie Author
%}%
%\author{Delta Author}
%\affiliation{%
% Authors' institution and/or address\\
% This line break forced with \textbackslash\textbackslash
%}%

\date{\today}

%\collaboration{CLEO Collaboration}%\noaffiliation

\begin{abstract}%
Low-energy $s$-wave scattering in weakly bound superfluid nuclei is strongly influenced by pairing correlations. In this work, we present an analytical study of low-energy $s$-wave quasiparticle scattering within the coordinate space Hartree-Fock-Bogoliubov framework using a schematic square-well model. Analytical expressions for the phase shift, elastic cross section, scattering length, and effective range are derived in a unified manner. We demonstrate that pairing correlations give rise to two distinct unitary limits, characterized by the divergence of the scattering length. One corresponds to the particle-like unitary limit, which persists even without pairing and is well described by the effective range expansion. The other is a pairing-induced hole-like unitary limit associated with quasiparticle resonances, leading to a breakdown of the effective range expansion. These results clarify the validity of the effective range expansion and highlight the essential role of resonance poles in describing low-energy $s$-wave quasiparticle scattering in superfluid nuclei.
\end{abstract}

%\keywords{Suggested keywords}%Use showkeys class option if keyword %display desired

\maketitle\relax\clearpage

%\section{\label{sec:level1}First-level heading:}

\section{Introduction}
The study of weakly bound nuclei is one of the central topics in modern nuclear physics~\cite{Ye2025}. Their exotic properties originate from the proximity of the Fermi surface to the particle continuum, which enhances coupling to unbound states. Pairing correlations, which play a dominant role in nuclear systems near the Fermi surface~\cite{BohrMottelson,RingSchuck,Dean2003,BrinkBroglia,BrogliaZelevinsky}, also have the continuum coupling~\cite{Dobaczewski1996,Meng2006}. As a result, they give rise to various phenomena such as neutron halo formation in drip-line nuclei~\cite{Tanihata1985,Hansen1987,Bertsch1991,Meng1996,Bennaceur2000,Barranco2001,Myo2002}.

Pairing correlations also have a strong impact on scattering phenomena at low energies. Within the coordinate space Hartree-Fock-Bogoliubov (HFB) framework, which enables a unified description of bound, continuum, and resonant states~\cite{Bulgac1980,Dobaczewski1984,Belyaev1987}, novel resonances, called {\it quasiparticle resonances}, have been predicted~\cite{Bulgac1980,Dobaczewski1984,Belyaev1987,Bennaceur1999,Fayans2000,Grasso2001,Hamamoto2003}. These resonances originate from the coupling between bound hole states and continuum states through pairing correlations and therefore provide direct insight into the role of pairing in open quantum systems.

In a series of previous studies, we have investigated the properties of quasiparticle resonances in weakly bound nuclei~\cite{Kobayashi2016,Kobayashi2020,Kobayashi2023}. It was shown that pairing correlations can generate pronounced resonance structures in the phase shift and elastic cross section even in $s$-wave scattering without a centrifugal barrier~\cite{Kobayashi2020}. Furthermore, these phenomena can be understood in terms of the underlying S-matrix pole structure associated with quasiparticle-quasihole symmetry~\cite{Kobayashi2020}. More recently, Ref.~\cite{Kobayashi2023} demonstrated that quasiparticle resonances manifest themselves as low-energy peaks in the decay spectrum of ${}^{21}$C.

It is interesting to clarify the low-energy scattering parameters, such as the scattering length and effective range, since these quantities are fundamental quantities characterizing low-energy $s$-wave scattering. They have long been used as indicators of the gross features of interaction potentials and the location of $s$-wave orbits and have played an important role in nuclear force studies~\cite{BohrMottelson,BlattWeisskopf}. The scattering length is also a key quantity in analyses of unbound subsystems, such as $(^{10}\mathrm{Li}+n)$, relevant to two-neutron halo nuclei like ${}^{11}$Li~\cite{Thompson1994,Jensen2004}. The scattering length estimated from experimental studies of ${}^{21}$C~\cite{Mosby2013} has been employed as an input in theoretical investigations of ${}^{22}$C~\cite{Pinilla2016,Shulgina2018,Singh2019}. More recently, scattering parameters have attracted renewed interest in discussions of universality, including Efimov states~\cite{Jensen2004,Braaten2006,Naidon2017}.

The purpose of the present study is to achieve an analytic and comprehensive understanding of low-energy $s$-wave quasiparticle scattering under the influence of pairing correlations. Using a schematic square-well model within the coordinate space HFB framework, we derive analytical expressions for the phase shift, elastic cross section, scattering length, effective range, and S-matrix. Based on these results, we demonstrate that pairing correlations give rise to two distinct unitary limits: a conventional particle-like unitary limit and a pairing-induced hole-like unitary limit associated with quasiparticle resonances. We further show that the validity of the effective range expansion differs sharply between the two unitary limits, reflecting their distinct physical origins.

This paper is organized as follows. In Sec.~2, we formulate analytically the phase shift, elastic cross section, scattering length, and effective range for $s$-wave quasiparticle scattering in the presence of pairing correlations. In Sec.~3, numerical results and discussions on pairing effects on these scattering parameters are presented. Finally, Sec.~4 summarizes the conclusions of this study.

\section{Theoretical framework}
\subsection{$s$-wave quasiparticle scattering with square-well potentials}
We formulate $s$-wave quasiparticle scattering in the presence of finite pairing correlations within the coordinate space HFB framework. As will be shown in the following subsection, the conventional $s$-wave single-particle scattering problem is recovered as a special limit of the present formulation, in which the pairing correlations are switched off. Throughout this paper we restrict ourselves to $s$-wave scattering and hence omit angular momentum indices. We neglect the Coulomb potential, and ignore spin and isospin degrees of freedom for simplicity.

In order to treat both pairing correlations and scattering states properly, we employ the coordinate space HFB theory. For a spherically symmetric system, the radial quasiparticle wave function $\varphi(r)=r^{-1}\left(u(r),~v(r) \right)^{T}$ satisfies the coordinate space HFB equation \cite{Dobaczewski1984,Matsuo2001},\begin{equation}
\left(
\begin{array}{ccc}
-\displaystyle\frac{\hbar^{2}}{2m}\frac{d^{2}}{dr^{2}}+V(r)-\lambda & \Delta (r) \\
\Delta (r) &\displaystyle\frac{\hbar^{2}}{2m}\frac{d^{2}}{dr^{2}}-V(r)+\lambda
\end{array}
\right)
\left(
\begin{array}{c}
u(r)\\
v(r)
\end{array}
\right)=E\left(
\begin{array}{c}
u(r)\\
v(r)
\end{array}
\right).
\end{equation}

Here, $u(r)$ and $v(r)$ denote the particle and hole components of the quasiparticle wave function, respectively. $V(r)$, $\lambda$, $\Delta(r)$, and $E$ represent the Hartree-Fock (HF) potential, the Fermi energy, the pairing potential, and the quasiparticle energy.  To allow for an analytic treatment of the scattering problem, both the HF potential and the pairing potential are taken to be square-wells,
\begin{equation}
V(r)=\left \{
\begin{array}{c}
-V_{0} \\
0
\end{array}
\right.,\quad \Delta(r)=\left \{
\begin{array}{c}
\Delta_{0} \\
0
\end{array}
\right.\quad
\begin{array}{l}
(r\le R,~V_{0}>0,~\Delta_{0}>0) \\
(r>R).
\end{array}
\end{equation}

Inside the well ($r\le R$), the quasiparticle wave function is given as a linear combination of two independent solutions,
\begin{equation}
\left(
\begin{array}{c}
u_{\mathrm{in}}(r) \\
v_{\mathrm{in}}(r)
\end{array}
\right)=A\left(
\begin{array}{c}
1 \\
\beta
\end{array}
\right)\sin k_{+}r+B\left(
\begin{array}{c}
\beta \\
1
\end{array}
\right)\sin k_{-}r~,
\end{equation}
where $A$ and $B$ are constants. The wave numbers $k_{\pm}$ and the coefficient $\beta$ are defined as
\begin{equation}
k_{\pm}=\sqrt{ \frac{2m}{\hbar^{2}} \left(V_{0}+\lambda \pm \sqrt{E^{2}-\Delta^{2}_{0}} \right) },\quad \beta=\frac{\Delta_{0}}{E+\sqrt{E^{2}-\Delta^{2}_{0}}}~.
\end{equation}

The quantities $k_{\pm}$ and $\beta$ are real for $E\ge\Delta_{0}$ and become complex for $E<\Delta_{0}$. In the region $E\ge\Delta_{0}$, one has $\beta\le 1$. Explicit expressions for the real and imaginary parts of $k_{\pm}$ and $\beta$ for $E<\Delta_{0}$, together with the corresponding quasiparticle wave functions, are
given in Appendix~A.

Outside the well ($r>R$), the quasiparticle wave function satisfying the scattering boundary condition is given by \cite{Bulgac1980,Dobaczewski1984,Belyaev1987},
\begin{equation}
\left(
\begin{array}{c}
u_{\mathrm{out}}(r)\\
v_{\mathrm{out}}(r)
\end{array}
\right)=\left(
\begin{array}{c}
C\left( \cos\delta\sin k_{1}r+\sin \delta\cos k_{1}r \right)\\
De^{-\kappa_{2}r}
\end{array}
\right),
\end{equation}
\begin{equation}
k_{1}=\sqrt{\frac{2m}{\hbar^{2}}(\lambda+E)},\quad k_{2}=\sqrt{\frac{2m}{\hbar^{2}}(\lambda-E)}=i\sqrt{\frac{2m}{\hbar^{2}}(E-\lambda)}=i\kappa_{2}~,
\end{equation}
\begin{equation}
C=\sqrt{\frac{2m}{\hbar^{2}\pi k_{1}}}~.
\end{equation}

Here $\delta$ is the phase shift of the particle component, while the hole component decays exponentially
outside the well, reflecting the fact that only the particle channel is open for scattering.

By imposing the matching conditions at $r=R$,
\begin{equation}
u_{\mathrm{in}}(R)=u_{\mathrm{out}}(R),\quad v_{\mathrm{in}}(R)=v_{\mathrm{out}}(R),
\end{equation}
\begin{equation}
\left.\frac{du_{\mathrm{in}}(r)}{dr}\right|_{r=R}=\left.\frac{du_{\mathrm{out}}(r)}{dr}\right|_{r=R}, \quad \left.\frac{dv_{\mathrm{in}}(r)}{dr}\right|_{r=R}=\left.\frac{dv_{\mathrm{out}}(r)}{dr}\right|_{r=R},
\end{equation}
the phase shift $\delta$ is obtained as
\begin{equation}
\tan\delta=-\frac{\mathcal{K}\sin k_{1}R-k_{1}\cos k_{1}R}{\mathcal{K}\cos k_{1}R+k_{1}\sin k_{1}R}~.
\end{equation}

The effects of the pairing potential $\Delta_{0}$ are fully incorporated into the quantity $\mathcal{K}$, whose explicit form is given in Appendix~B. The elastic cross section is given, as in the single-particle case, by $\sigma=(4\pi/k_{1}^{2})\sin^{2}\delta$.

The effective range expansion provides a useful parametrization of low-energy $s$-wave scattering~\cite{BlattWeisskopf,Messiah,Taylor}. We will examine  its validity in quasiparticle scattering with pairing correlations.  The scattering length $a$ and the effective range $r_{\mathrm{eff}}$ are defined as the coefficients of the constant and $k_{1}^{2}$ terms in the low-$k_{1}$ (i.e., $E\sim -\lambda$) expansion of $k_{1}\cot\delta$,
\begin{equation}
k_{1}\cot \delta=-\frac{1}{a}+\frac{1}{2}r_{\mathrm{eff}}k^{2}_{1}+\mathcal{O}\left( k^{4}_{1} \right).
\end{equation}

For the square-well model, the scattering length $a$ and the effective range $r_{\mathrm{eff}}$ for quasiparticle scattering can also be obtained analytically as
\begin{equation}
a=R-\frac{1}{\tilde{\mathcal{K}}},\quad r_{\mathrm{eff}}=2R\left( 1-\frac{R}{a}+\frac{R^{2}}{3a^{2}} \right)+2\mathcal{R}\left( 1-\frac{R}{a} \right)^{2},
\end{equation}
where $\mathcal{K}$ has been expanded with respect to $k_{1}$ as
\begin{equation}
\mathcal{K}=\tilde{\mathcal{K}}+\mathcal{R} k^{2}_{1}, \quad \lim_{k_{1}\to0}\mathcal{K}=\tilde{\mathcal{K}}~.
\end{equation}

Here and in the following, a tilde denotes the zero-energy limit $k_{1}\to0$ (i.e., $E\to-\lambda$) of the quantity beneath it, that is, the constant term in its expansion in powers of $k^{2}_{1}$. Explicit expressions for $\tilde{\mathcal{K}}$ and $\mathcal{R}$ are presented in Appendix~C. From Eq.~(12), one finds that at the unitary limits $a=\pm\infty$ the effective range behaves as $r_{\mathrm{eff}}= 2R+2\mathcal{R}$.

As in the single-particle scattering problem, the S-matrix can be defined in terms of the phase shift for complex $k_{1}$ as
\begin{equation}
S\left( k_{1} \right)=\frac{\cot\delta(k_{1})+i}{\cot\delta(k_{1})-i}~.
\end{equation}

A crucial difference from the single-particle case is that two wave numbers, $k_{1}$ and $k_{2}$, must be considered in the quasiparticle scattering problem. Consequently the complex $E$-plane consists of four Riemann sheets according to the signs of $\mathrm{Im}(k_{1})$ and $\mathrm{Im}(k_{2})$. Following Ref.~\cite{Kobayashi2020}, we classify the four Riemann sheets as summarized in Table~1. Further details on the properties of the four poles can be found in Ref.~\cite{Kobayashi2020}.

\begin{table}[ht]
\caption{The Riemann sheets classified by $\mathrm{Im}(k_{1})$ and $\mathrm{Im}(k_{2})$.}
\begin{center}
\begin{tabular}{cc} \hline
$E^{(1)}$-sheet: & $~\mathrm{Im}(k_{1})>0,~\mathrm{Im}(k_{2})>0$ \\
$E^{(2)}$-sheet: & $~\mathrm{Im}(k_{1})<0,~\mathrm{Im}(k_{2})>0$ \\
$E^{(3)}$-sheet: & $~\mathrm{Im}(k_{1})>0,~\mathrm{Im}(k_{2})<0$ \\
$E^{(4)}$-sheet: & $~\mathrm{Im}(k_{1})<0,~\mathrm{Im}(k_{2})<0$ \\ \hline
\end{tabular}
\end{center}
\end{table}

The S-matrix has four poles. There are two poles corresponding to those appearing in potential scattering without pairing. We denote them $a$ and $b$ (see Sec.~2.2). The quasiparticle-quasihole symmetry inherent in the HFB theory,
\begin{equation}
\left(
\begin{array}{c}
u(r,-E)\\
v(r,~-E)
\end{array}
\right)=\left(
\begin{array}{c}
-v(r,E)\\
u(r,~E)
\end{array}
\right),
\end{equation}
generates an additional pair of poles, $\bar{a}$ and $\bar{b}$, leading to a total of four poles~\cite{Kobayashi2020}. The poles $a$ and $\bar{a}$ satisfy the relation
\begin{equation}
\left( \bar{k}_{1,\bar{a}},~\bar{k}_{2,\bar{a}},~\bar{E}_{\bar{a}} \right)=\left( \bar{k}_{2,a},~\bar{k}_{1,a},~-\bar{E}_{a} \right),
\end{equation}
and an analogous relation holds between poles $b$ and $\bar{b}$. Consequently, the motion of poles in the complex $k_{2}$-plane can be inferred from their behavior in the complex $k_{1}$-plane.

\subsection{$s$-wave single-particle scattering as the $\Delta_{0}\to0$ limit of quasiparticle scattering}
The conventional $s$-wave single-particle scattering problem is recovered from the quasiparticle formulation of Sec.~2.1 by switching off the pairing potential, $\Delta_{0}\to0$. In this limit, $\beta\to0$, so that the particle and hole components $u(r)$ and $v(r)$ in Eqs.~(3) and (5) decouple completely. Since the quasiparticle and single-particle energies are related by $e=E+\lambda$ in this limit, the wave numbers
reduce as
\begin{equation}
\lim_{\Delta_{0}\to0}k_{+}=k_{\mathrm{in}}=\sqrt{\frac{2m}{\hbar^{2}}(V_{0}+e)}, \quad \lim_{\Delta_{0}\to0}k_{1}=k_{\mathrm{out}}=\sqrt{\frac{2m}{\hbar^{2}}e}.
\end{equation}

The hole-sector quantities ($k_{-}$, $\kappa_{2}$, $v(r)$) drop out entirely since they carry no scattering information once $\beta=0$. The remaining particle component, $u_{\mathrm{in}}(r)\to A\sin k_{\mathrm{in}}r$ and $u_{\mathrm{out}}(r)\to C(\cos\delta\sin k_{\mathrm{out}}r+\sin\delta\cos k_{\mathrm{out}}r)$, is precisely the ordinary single-particle scattering wave function for the square-well.

Applying the same limit to the quantity $\mathcal{K}$ in Eq.~(10) gives $\displaystyle\lim_{\Delta_{0}\to0}\mathcal{K}=k_{\mathrm{in}}\cot(k_{\mathrm{in}}R)$, so that the quasiparticle phase shift, scattering length, effective range, and S-matrix of Sec.~2.1 reduce, term by term, to their familiar single-particle forms. The phase shift~(10) becomes
\begin{equation}
\lim_{\Delta_{0}\to0}\tan\delta=-\frac{k_{\mathrm{in}}\cot k_{\mathrm{in}}R \sin k_{\mathrm{out}}R-k_{\mathrm{out}}\cos k_{\mathrm{out}}R}{k_{\mathrm{in}}\cot k_{\mathrm{in}}R\cos k_{\mathrm{out}}R+k_{\mathrm{out}}\sin k_{\mathrm{out}}R}~,
\end{equation}
and the elastic cross section is
\begin{equation}
\lim_{\Delta_{0}\to0}\sigma=\frac{4\pi}{k_{\mathrm{out}}^{2}}\sin^{2}\left( \lim_{\Delta_{0}\to0}\delta \right).
\end{equation}

$\displaystyle\lim_{\Delta_{0}\to0}a$ and $\displaystyle\lim_{\Delta_{0}\to0}r_{\mathrm{eff}}$ are obtained from
Eq.~(12) by taking the $\Delta_{0}\to0$ limit of $\tilde{\mathcal{K}}$ and $\mathcal{R}$. In this limit, $\tilde{\mathcal{K}}$ reduces to
\begin{equation}
\lim_{\Delta_{0}\to0}\tilde{\mathcal{K}}=\tilde{k}_{\mathrm{in}}\cot(\tilde{k}_{\mathrm{in}}R),\quad \tilde{k}_{\mathrm{in}}=\lim_{k_{\mathrm{out}}\to0}k_{\mathrm{in}}=\sqrt{\frac{2m}{\hbar^2}V_{0}}~,
\end{equation}
so that the scattering length reduces accordingly to
\begin{equation}
\lim_{\Delta_{0}\to0}a=R-\frac{1}{\tilde{k}_{\mathrm{in}}\cot \tilde{k}_{\mathrm{in}}R}
\equiv a_0.
\end{equation}

Likewise, $\mathcal{R}$ reduces, in the same limit, to
\begin{equation}
\lim_{\Delta_{0}\to0}\mathcal{R}=-\frac{R}{2}-\frac{a_0}{2\tilde{k}^{2}_{\mathrm{in}}\left( R-a_0 \right)^{2}},
\end{equation}
so that the effective range reduces to
\begin{equation}
\lim_{\Delta_{0}\to0}r_{\mathrm{eff}}=R-\frac{R^{3}}{3 a_0^{2}}-\frac{1}{\tilde{k}^{2}_{\mathrm{in}}a_0} \equiv r_{\mathrm{eff},0}.
\end{equation}

From Eq.~(23), $\displaystyle\lim_{\Delta_{0}\to0} r_{\rm eff}=R$ at the unitary limit
$a_0=\pm\infty$, whereas $r_{{\rm eff},0}\to-\infty$ at $ a_0=0$. These are the well-known relations of the
single-particle scattering problem.

The S-matrix~(14) reduces to $S(k_{\mathrm{out}})=[\cot\delta(k_{\mathrm{out}})+i]/[\cot\delta(k_{\mathrm{out}})-i]$, whose poles $\bar{k}_{\mathrm{out},a}$ classify single-particle bound, virtual, and resonances. Denoting a pole by $\bar{k}_{\mathrm{out},a}$, a bound state corresponds to $\mathrm{Re}(\bar{k}_{\mathrm{out},a})=0$ and $\mathrm{Im}(\bar{k}_{\mathrm{out},a})>0$, a virtual state to $\mathrm{Re}(\bar{k}_{\mathrm{out},a})=0$ and $\mathrm{Im}(\bar{k}_{\mathrm{out},a})<0$, and a resonance to $\mathrm{Re}(\bar{k}_{\mathrm{out},a})>0$ and $\mathrm{Im}(\bar{k}_{\mathrm{out},a})<0$.

Each pole $\bar{k}_{\mathrm{out},a}$ is accompanied by a conjugate pole $\bar{k}_{\mathrm{out},b}$. When pole $a$ corresponds to a resonance, pole $b$ is referred to as an anti-resonance, and their locations satisfy the Schwarz reflection principle, $\left( \bar{k}_{\mathrm{out},b},~\bar{e}_{b} \right)=\left(-\bar{k}^{\ast}_{\mathrm{out},a},~\bar{e}^{\ast}_{a} \right)$. The additional poles $\bar{a},~\bar{b}$ generated by the quasiparticle-quasihole symmetry~(15) are absent here.

\section{Results and Discussion}
\subsection{$V_{0}$ dependence of $s$-wave single-particle scattering}
As a reference for comparison, we first present the calculated results for $s$-wave single-particle scattering. The radius of the square-well potential is fixed to $R=5.0$ fm throughout this subsection. Figure~1 shows the energy dependence of the phase shift $\delta$ and the elastic cross section $\sigma$ for several values of the potential depth $V_{0}$. Near zero energy ($e\sim 0$), the phase shift $\delta$ exhibits a decreasing trend for $V_{0}=55.0-52.0$ MeV. As the potential becomes shallower and reaches $V_{0}=51.0$ MeV, the behavior of $\delta$ near $e \sim 0$ changes to an increasing trend. These features indicate that the weakly bound $s$ orbit changes from a bound state to an unbound (virtual) state in the region $V_{0}=52.0-51.0$ MeV. Correspondingly, the elastic cross section $\sigma$ shows a sharp enhancement near zero energy for $V_{0}=52.0-51.0$ MeV.

\begin{figure}[ht]
\begin{center}
\includegraphics[width=80mm]{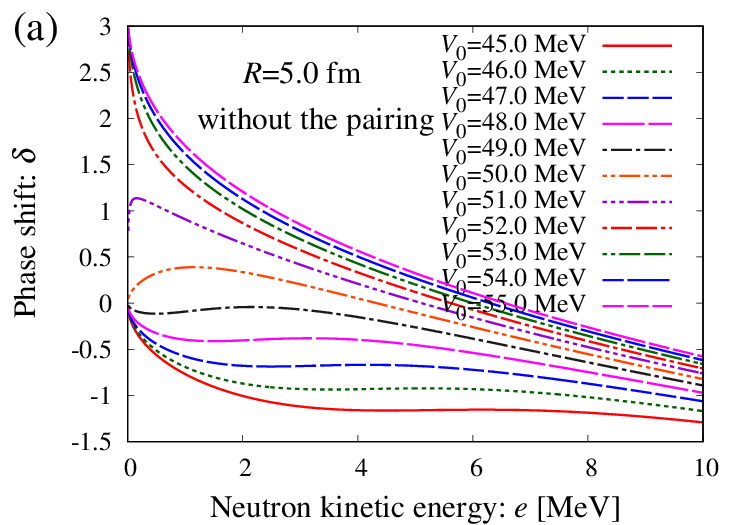}
\includegraphics[width=80mm]{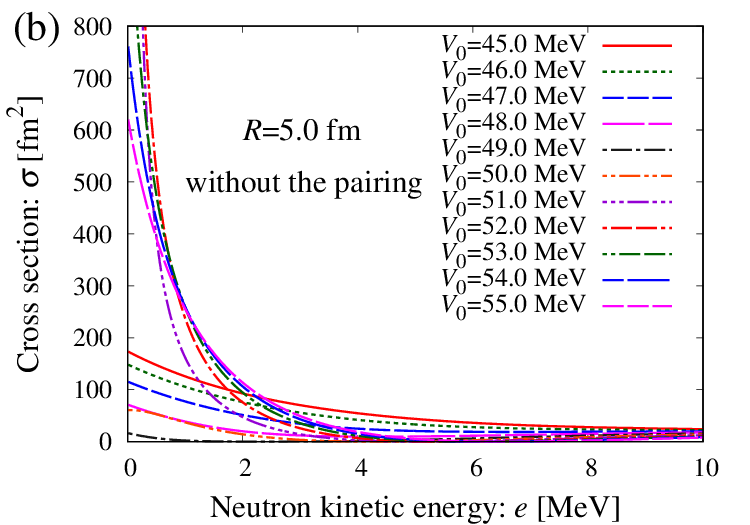}
\caption{(a) Phase shift $\delta(e)$ and (b) cross section $\sigma(e)$ of $s$-wave elastic scattering for various values of the potential depth $V_{0}$ without the pairing ($\Delta_{0}=0$ MeV). The horizontal axis is the kinetic energy of scattered neutron $e$.}
\end{center}
\end{figure}

Figure~2 displays the dependence of the scattering length $a$ and the effective range $r_{\mathrm{eff}}$ on the potential depth $V_{0}$. The scattering length increases rapidly as $V_{0}$ approaches $\sim 51$ MeV and diverges to $\pm\infty$ at $V_{0}=51.156$ MeV, which corresponds to the unitary limit. As $V_{0}$ is further decreased beyond the unitary limit, the scattering length approaches zero and becomes exactly $a=0$ at $V_{0}=49.493$ MeV. The behaviors shown in Fig.~2(a) and Fig.~1(a) can be understood from the relation $\delta \sim \arctan(-a k_{\mathrm{out}}) \sim -a k_{\mathrm{out}}$, which follows from the effective range expansion.

As shown in Fig.~2(b), the effective range takes the value $r_{\mathrm{eff}}=R=5.0$ fm at the unitary limit ($a=\pm\infty$ with $V_{0}=51.156$ MeV). When $V_{0}\gtrsim51$ MeV, namely when the $s$ orbit near the binding threshold ($e=0$) remains bound, the effective range satisfies $r_{\mathrm{eff}}\sim R$. In contrast, at $a=0$ ($V_{0}=49.493$ MeV), the effective range diverges to $r_{\mathrm{eff}}=-\infty$. All of these features are consistently understood from Eq.~(23).

\begin{figure}[ht]
\begin{center}
\includegraphics[width=80mm]{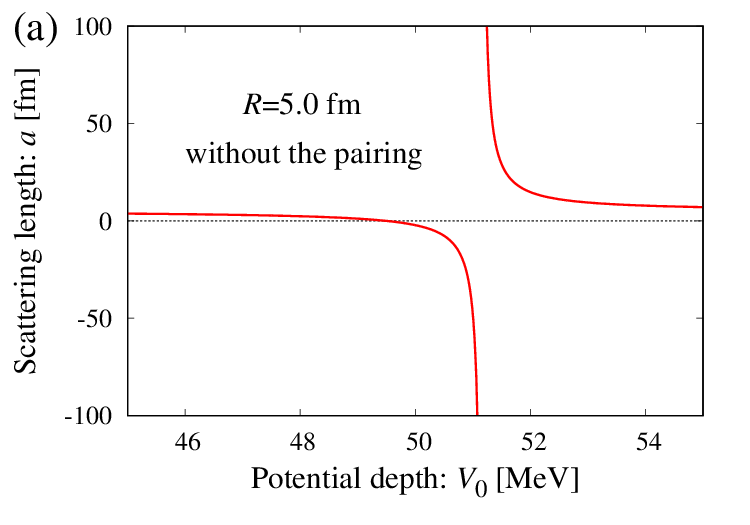}
\includegraphics[width=80mm]{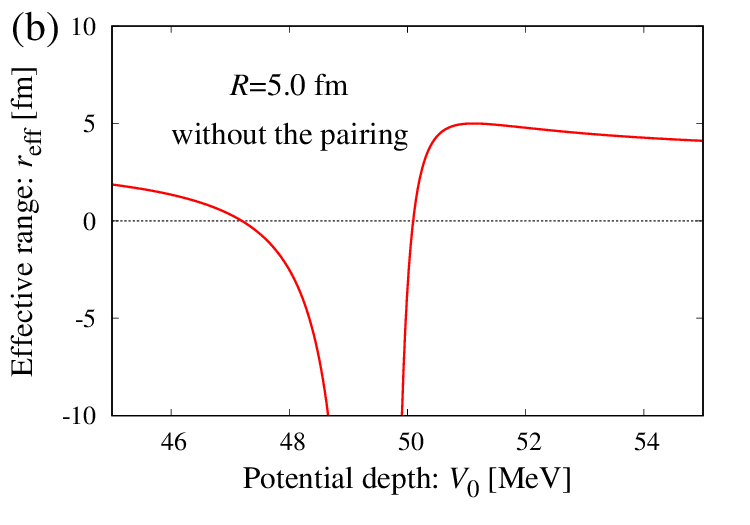}
\caption{Dependence on the potential depth $V_{0}$ of (a) scattering length $a(V_{0})$ and (b) effective range $r_{\mathrm{eff}}(V_{0})$ of $s$-wave elastic scattering without the pairing ($\Delta_{0}=0$ MeV).}
\end{center}
\end{figure}

Next, we examine the behaviors of $\delta$ and $k_{\mathrm{out}}\cot\delta$ in the vicinity of $a=\pm\infty$ and $a=0$. Figure~3 shows the low-energy behavior of $\delta$ and $k_{\mathrm{out}}\cot\delta$ around the unitary limit ($V_{0}\approx 51.1$ MeV). As seen in Fig.~3(a), the slope of $\delta$ at $e\sim 0$ ($k_{\mathrm{out}}\sim 0$) changes from $+\infty$ to $-\infty$ as $V_{0}$ varies from $51.1$ to $51.2$ MeV. Figure~3(b) shows that the intercept of $k_{\mathrm{out}}\cot\delta$, corresponding to $-1/a$, changes from $+0$ to $-0$ in the same range of $V_{0}$. In addition, $k_{\mathrm{out}}\cot\delta$ exhibits an approximately linear dependence on $k_{\mathrm{out}}^{2}$, with a slope corresponding to $r_{\mathrm{eff}}/2\sim 2.5$.

\begin{figure}[ht]
\begin{center}
\includegraphics[width=80mm]{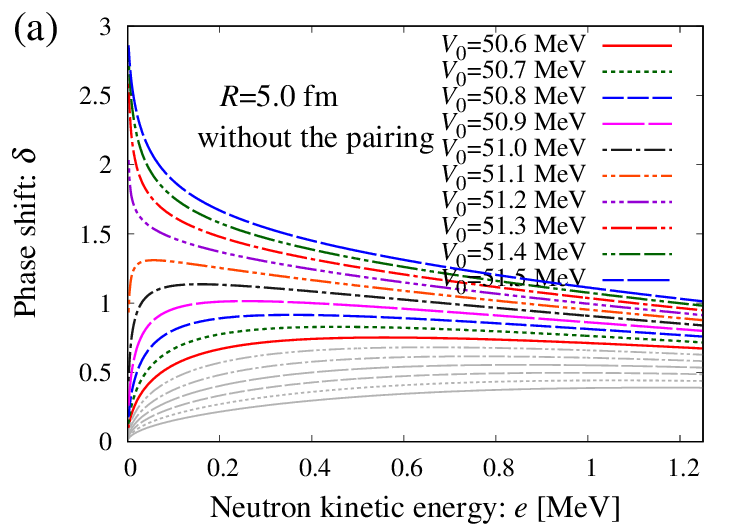}
\includegraphics[width=80mm]{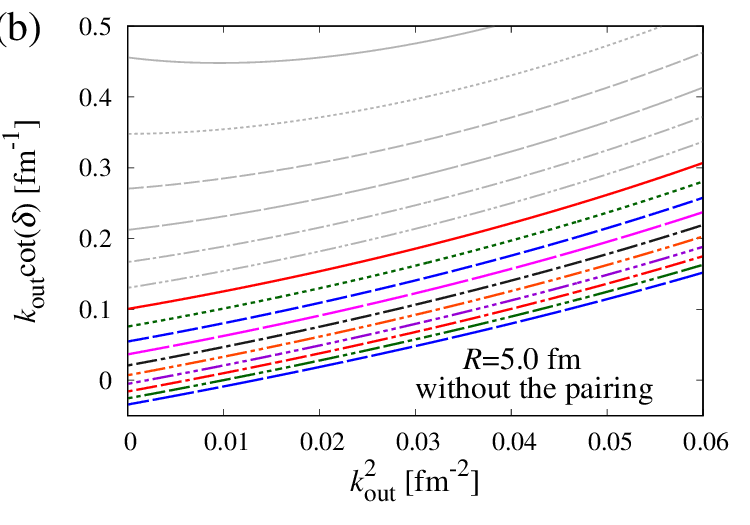}
\caption{(a) $\delta$ and (b) $k_{\mathrm{out}}\cot\delta$ of $s$-wave elastic scattering for $V_{0}=50.6-51.5$ MeV without the pairing ($\Delta_{0}=0$ MeV). For comparison with Fig.~4, the results for $V_{0}=50.0-50.5$ MeV are shown by the gray lines, with the gray solid line corresponding to the result for $V_{0}=50.0$ MeV. The legend is common to both figures.}
\end{center}
\end{figure}

Figure~4 presents the results near $V_{0}=49.5$ MeV, corresponding to the vicinity of $a=0$. As shown in Fig.~4(a), the slope of $\delta$ at $e\sim 0$ becomes zero for $V_{0}=49.4-49.5$ MeV. Figure~4(b) shows that the intercept of $k_{\mathrm{out}}\cot\delta$ changes from $-\infty$ to $+\infty$ in this region, reflecting the change of the scattering length from $+0$ to $-0$. Under these conditions, the slope of $k_{\mathrm{out}}\cot\delta$ in the low-energy region, namely $r_{\mathrm{eff}}/2$, is found to take a large negative value.

\begin{figure}[ht]
\begin{center}
\includegraphics[width=80mm]{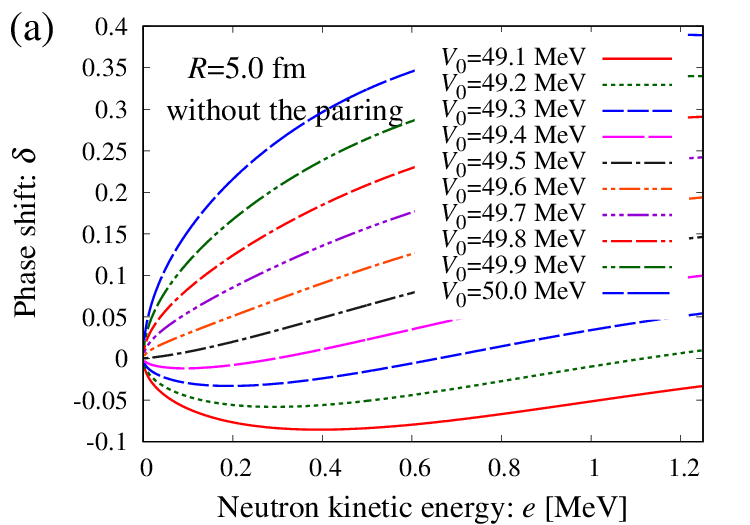}
\includegraphics[width=80mm]{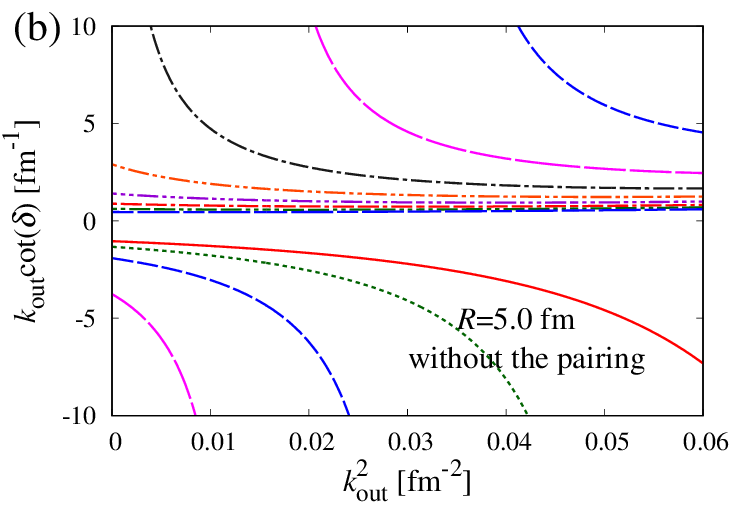}
\caption{The same as Fig. 3, but for $V_{0}=49.1-50.0$ MeV. The legend is common to both figures.}
\end{center}
\end{figure}

The results shown in Figs.~3 and~4 thus reconfirm the well-known properties of the scattering length and the effective range through the low-energy effective range expansion.

Finally, we examine the dependence of the S-matrix poles on the potential depth $V_{0}$. Figure~5 shows the trajectory of the S-matrix pole $a$ as $V_{0}$ is varied, while the trajectory of the conjugate pole $b$ is omitted for simplicity. For deep potentials ($V_{0}\gtrsim 52$ MeV), the pole is located on the negative real axis of the $e^{(1)}$-sheet ($\mathrm{Im}(k_{\mathrm{out}})>0$). As the potential becomes shallower, the pole approaches the origin and moves to the negative real axis of the $e^{(2)}$-sheet ($\mathrm{Im}(k_{\mathrm{out}})<0$) at $V_{0}=51.156$ MeV, corresponding to the unitary limit ($a=\pm\infty$). With further reduction of $V_{0}$, the pole acquires an imaginary part for $V_{0}\le 50.322$ MeV and exhibits resonance-like behavior. However, since the imaginary part is large compared to the real part, no pronounced peak appears in the elastic cross section. As $V_{0}$ is decreased further, the pole crosses the imaginary axis ($\mathrm{Re}(e)=0$) at $V_{0}=49.469$ MeV.

\begin{figure}[ht]
\begin{center}
\includegraphics[width=80mm]{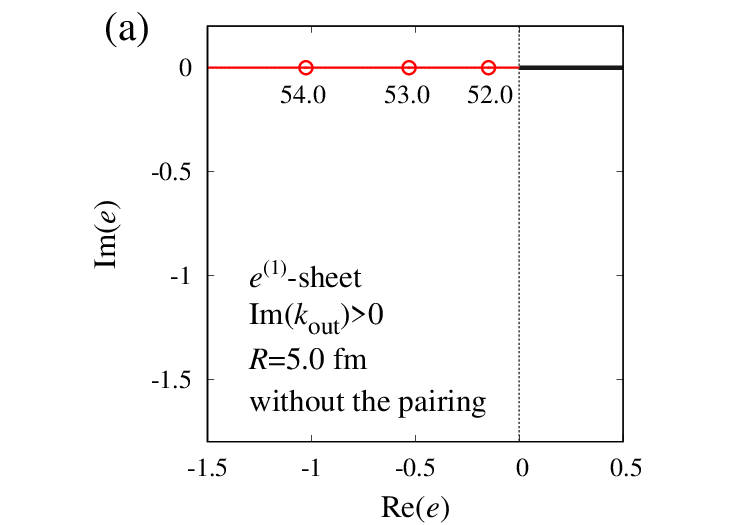}
\includegraphics[width=80mm]{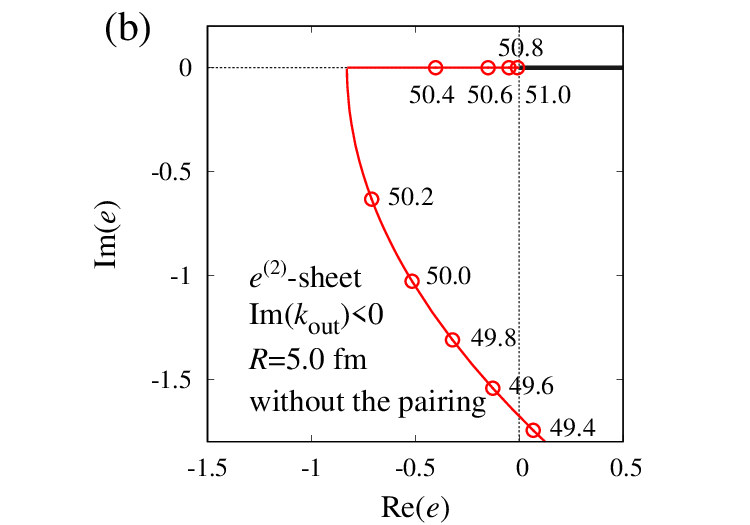}
\caption{Pole trajectory on the Riemann sheets of the complex $e$-plane, where the potential depth is varied in the range $V_{0}=49.0-55.0$ MeV while the potential width is fixed to $R=5.0$ fm. The trajectory is plotted separately on the two Riemann sheets, and marked with symbols at intervals of 1.0 MeV in $V_{0}$ for panel (a) and 0.2 MeV for panel (b). The thick lines on the real $e$-axis are the branch cuts.}
\end{center}
\end{figure}

\subsection{Pairing effects on $s$-wave quasiparticle scattering}
In this subsection, we discuss the effects of pairing correlations on $s$-wave quasiparticle scattering ($\Delta_{0}\neq 0$), with a direct comparison to the single-particle scattering problem reviewed in Sec.~3.1. In our previous studies~\cite{Kobayashi2016,Kobayashi2020}, pairing effects on the phase shift, elastic scattering cross section, and S-matrix poles were investigated. The consistency with those earlier results is also examined here.

The radius of the square-well potential is fixed to $R=5.0$ fm, as in the previous subsection, while the potential depth $V_{0}$ is varied. The Fermi energy is set to $\lambda=-0.5$ MeV, corresponding to a weakly bound (drip-line) nucleus. Quasiparticle scattering states are defined for $E>|\lambda|$~\cite{Bulgac1980,Dobaczewski1984,Belyaev1987}.

Figure~6 shows the energy dependence of the phase shift $\delta$ and the elastic cross section $\sigma$ for several values of $V_{0}$ in the presence of pairing correlations. The pairing strength is fixed to $\Delta_{0}=1.0$ MeV. In contrast to the single-particle case shown in Fig.~1, the overall behavior in Fig.~6 clearly demonstrates that pairing correlations strongly affect low-energy $s$-wave scattering. In particular, the pronounced resonance-like structures observed for $V_{0}=54.0$ and $55.0$ MeV are characteristic. These resonances correspond to hole-like quasiparticle resonances originating from bound $s$-orbit (hole) states~\cite{Kobayashi2016,Kobayashi2020}.

\begin{figure}[ht]
\begin{center}
\includegraphics[width=80mm]{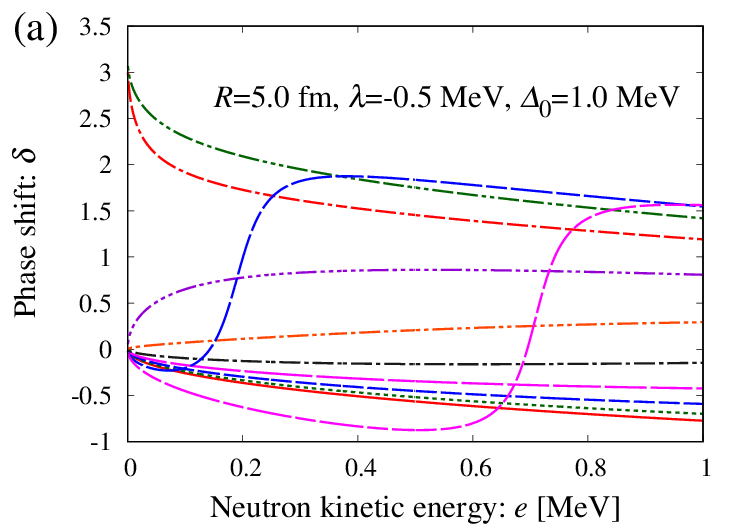}
\includegraphics[width=80mm]{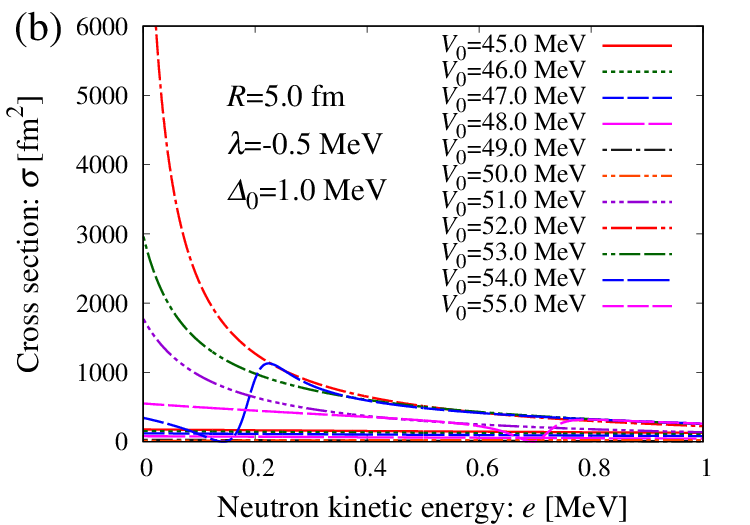}
\caption{(a) Phase shift $\delta(e)$ and (b) cross section $\sigma(e)$ of $s$-wave elastic quasiparticle scattering for various values of the potential depth $V_{0}$ with $\Delta_{0}=1.0$ MeV. The horizontal axis is the kinetic energy of scattered neutron $e$. The legend is common to both figures.}
\end{center}
\end{figure}

Figure~7 displays the dependence of the scattering length $a$ and the effective range $r_{\mathrm{eff}}$ on the potential depth $V_{0}$ under finite pairing correlations. Results for several values of the pairing strength $\Delta_{0}$ are shown. For comparison, the result for $\Delta_{0}=0.0$ MeV (single-particle case; Fig.~2) is also included.

\begin{figure}[ht]
\begin{center}
\includegraphics[width=80mm]{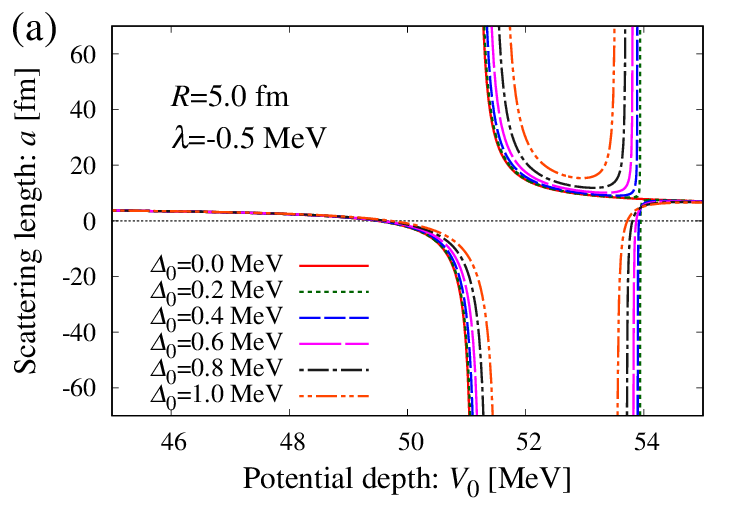}
\includegraphics[width=80mm]{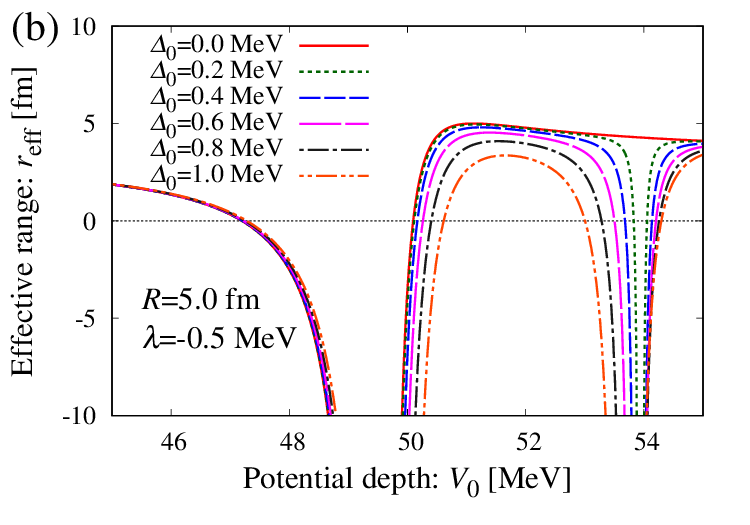}
\caption{Dependence on the potential depth $V_{0}$ of (a) scattering length $a(V_{0})$ and (b) effective range $r_{\mathrm{eff}}(V_{0})$ of $s$-wave elastic quasiparticle scattering with various $\Delta_{0}$.}
\end{center}
\end{figure}

By comparing the results for $\Delta_{0}=0.0$ MeV and $\Delta_{0}\neq 0.0$ MeV, one clearly observes that pairing correlations significantly modify both the scattering length $a$ and the effective range $r_{\mathrm{eff}}$. In particular, for $\Delta_{0}\neq 0.0$ MeV, two distinct unitary limits with $a=\pm\infty$ appear. The unitary limit around $V_{0}\sim 51.0$ MeV is also present in the $\Delta_{0}=0.0$ MeV case, whereas the unitary limit around $V_{0}\sim 53.5$ MeV emerges only when pairing correlations are present. For later discussion, we refer to the former as the {\it particle-like unitary limit} $U_{\mathrm{p}}$ and the latter as the {\it hole-like unitary limit} $U_{\mathrm{h}}$. In addition, Fig.~7 shows that the scattering length vanishes ($a=0$) around $V_{0}\sim 49.7$ MeV and $V_{0}\sim 53.7$ MeV, where the effective range diverges to $r_{\mathrm{eff}}=-\infty$.

\subsection{Properties of the unitary limits $U_{\mathrm{p}}$ and $U_{\mathrm{h}}$ under the pairing}
The most striking consequence of pairing correlations is the existence of two distinct unitary limits: the particle-like unitary limit $U_{\mathrm{p}}$ and the hole-like unitary limit $U_{\mathrm{h}}$. Although both are characterized by a divergent scattering length, their physical origins and implications for low-energy scattering are fundamentally different.

In the following, we examine in detail the characteristics of the phase shift, scattering length, and effective range at the particle-like and hole-like unitary limits, $U_{\mathrm{p}}$ and $U_{\mathrm{h}}$, respectively.

\subsubsection{Particle-like unitary limit $U_{\mathrm{p}}$}
To examine the properties of the scattering length $a$ and the effective range $r_{\mathrm{eff}}$ at the particle-like unitary limit $U_{\mathrm{p}}$, we investigate the low-energy behavior of the phase shift $\delta$ and $k_{1}\cot\delta$. Figure~8 shows $\delta$ and $k_{1}\cot\delta$ in the low-energy region in the vicinity of $U_{\mathrm{p}}$ ($V_{0}\sim 51.5$ MeV).

\begin{figure}[ht]
\begin{center}
\includegraphics[width=80mm]{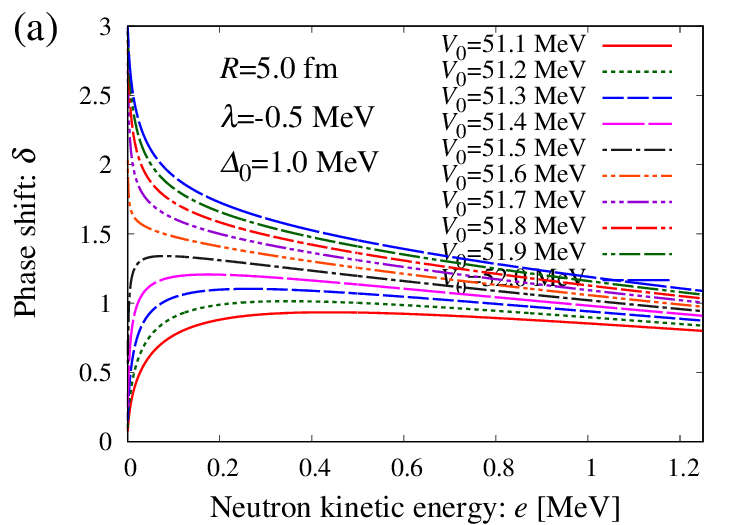}
\includegraphics[width=80mm]{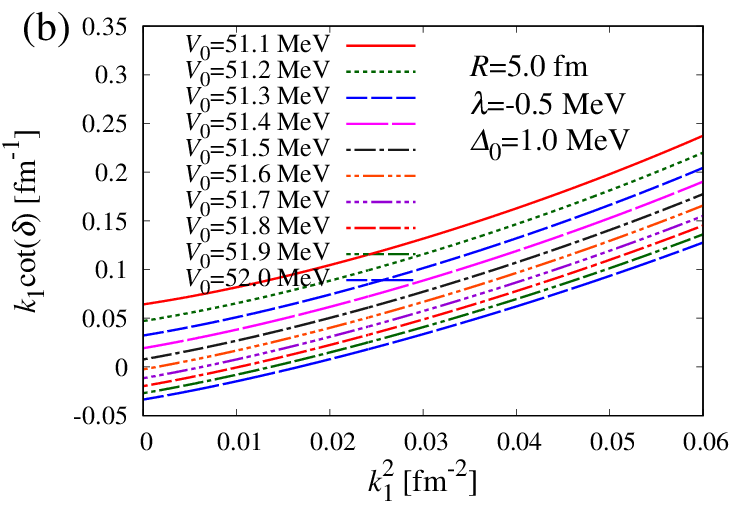}
\caption{(a) $\delta$ and (b) $k_{1}\cot\delta$ of $s$-wave elastic quasiparticle scattering for $V_{0}=51.1-52.0$ MeV with $\Delta_{0}=1.0$ MeV.}
\end{center}
\end{figure}

The behaviors of $\delta$ and $k_{1}\cot\delta$ in Fig.~8 are found to be similar to those observed in the single-particle elastic scattering problem (Fig.~3). As shown in Fig.~8(a), $\delta$ near $E\sim-\lambda$ ($k_{1}\sim 0$) changes from an increasing trend to a decreasing one as $V_{0}$ varies from $51.5$ to $51.6$ MeV. Correspondingly, the intercept of $k_{1}\cot\delta$, namely $-1/a$, changes from $+0$ to $-0$ in the same range of $V_{0}$, indicating that the scattering length $a$ diverges from $-\infty$ to $+\infty$. This behavior near $E\sim -\lambda$ is also understood from the relation $\delta \sim -a k_{1}$.

In comparison with Fig.~3(b), Fig.~8(b) shows that the slope of $k_{1}\cot\delta$ is smaller than the value $r_{\rm eff}/2\simeq2.5$ obtained without pairing. This indicates that pairing correlations reduce the effective range $r_{\mathrm{eff}}$ at $U_{\mathrm{p}}$. Table~2 lists the values of the potential depth $V_{0}$ corresponding to $U_{\mathrm{p}}$ and the effective range $r_{\mathrm{eff}}$ for several values of the pairing strength $\Delta_{0}$. As shown in Table~2, the effective range at $U_{\mathrm{p}}$ decreases systematically with increasing $\Delta_{0}$.

\begin{table}[ht]
\caption{The values for potential depth $V_{0}$ and effective range $r_{\mathrm{eff}}$ of $s$-wave elastic quasiparticle scattering at the particle-like unitary limits $U_{\mathrm{p}}$.}
\begin{center}
\begin{tabular}{ccc} \hline
$\Delta_{0}$ [MeV] & $V_{0}$ [MeV] & $r_{\mathrm{eff}}$ [$\mathrm{fm}$]\\ \hline
0.0 & 51.156 & 5.000\\
0.2 & 51.171 & 4.952\\
0.4 & 51.214 & 4.801\\
0.6 & 51.291 & 4.526\\
0.8 & 51.406 & 4.080\\
1.0 & 51.575 & 3.353\\ \hline
\end{tabular}
\end{center}
\end{table}

Next, we examine the behavior of the S-matrix pole in the vicinity of $U_{\mathrm{p}}$. Figure~9 presents the $V_{0}$ dependence of the pole $a$ around $U_{\mathrm{p}}$ for $\Delta_{0}=1.0$ MeV. As $V_{0}$ decreases, the pole initially moves along the real axis on the $E^{(1)}$-sheet (with $\mathrm{Re}(E)<-\lambda$) toward negative energies, turns around at $V_{0}=52.783$ MeV, and then starts moving in the positive direction. The pole subsequently passes through the threshold $E=(-\lambda,0)$ at $V_{0}=51.575$ MeV and moves to the $E^{(2)}$-sheet, which corresponds to the particle-like unitary limit $U_{\mathrm{p}}$. The pole trajectory near the threshold as $V_{0}$ decreases is qualitatively similar to that observed in the single-particle scattering problem (Fig.~5).

\begin{figure}[ht]
\begin{center}
\includegraphics[width=80mm]{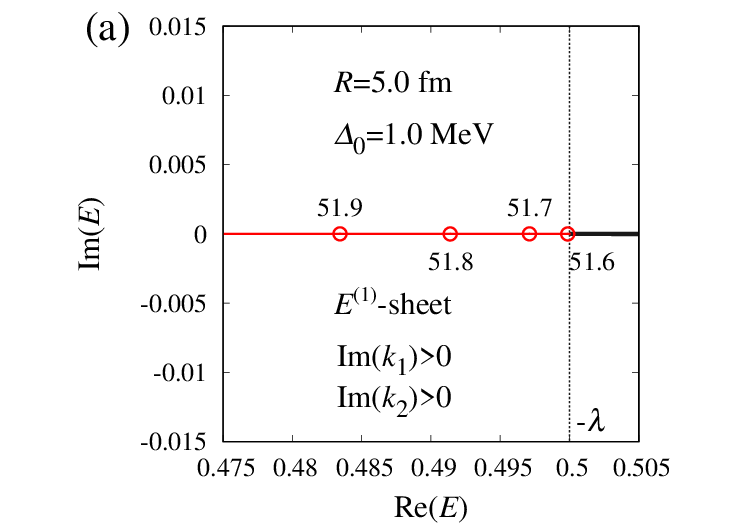}
\includegraphics[width=80mm]{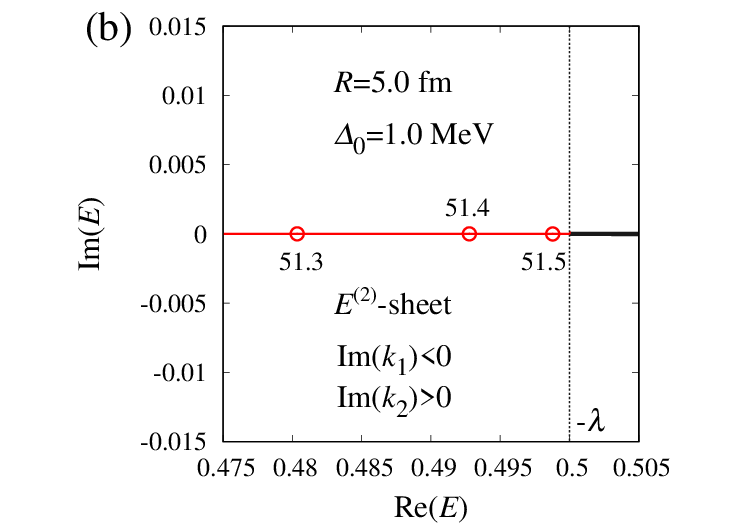}
\caption{Pole trajectory on the Riemann sheets of the complex $E$-plane, where the potential depth is varied in the range $V_{0}=51.2-52.0$ MeV while the pairing strength is fixed to $\Delta_{0}=1.0$ MeV. The trajectory is plotted separately on the two Riemann sheets, and marked with symbols for every 0.1 MeV interval in $V_{0}$. The thick lines on the real $E$-axis are the branch cuts.}
\end{center}
\end{figure}

\subsubsection{Hole-like unitary limit $U_{\mathrm{h}}$}
Figure~10 shows the phase shift $\delta$ and $k_{1}\cot\delta$ in the vicinity of the hole-like unitary limit $U_{\mathrm{h}}$ ($V_{0}\sim 53.5$ MeV). As seen in Fig.~10(a), pronounced resonance-like structures appear for $V_{0}=53.6-54.0$ MeV. These structures correspond to the hole-like quasiparticle resonances discussed above. For $V_{0}=53.8-54.0$ MeV, the background contribution leads to zero crossings of the phase shift, $\delta=0$, at finite energies $e>0$. Near $e\sim 0$, the behavior of $\delta$ changes from an increasing trend to a decreasing one as $V_{0}$ varies from $53.5$ to $53.6$ MeV, accompanied by quasiparticle resonances. This resonance-assisted change is in marked contrast to the behavior observed at the particle-like unitary limit $U_{\mathrm{p}}$ (Fig.~8(a)).

\begin{figure}[ht]
\begin{center}
\includegraphics[width=80mm]{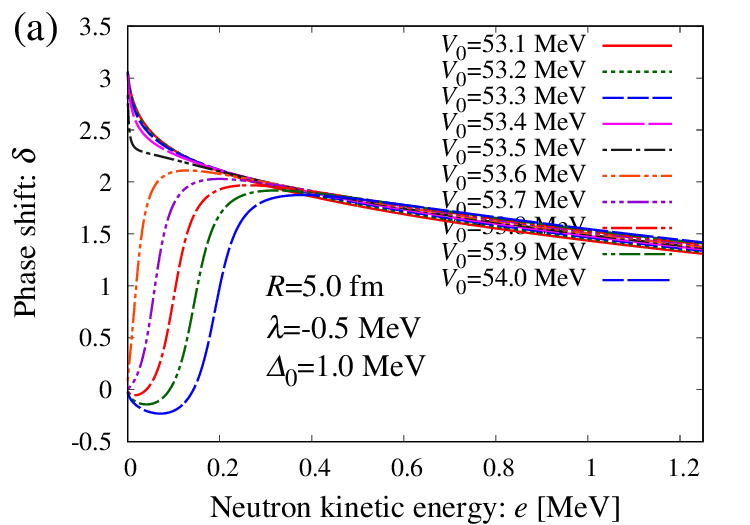}
\includegraphics[width=80mm]{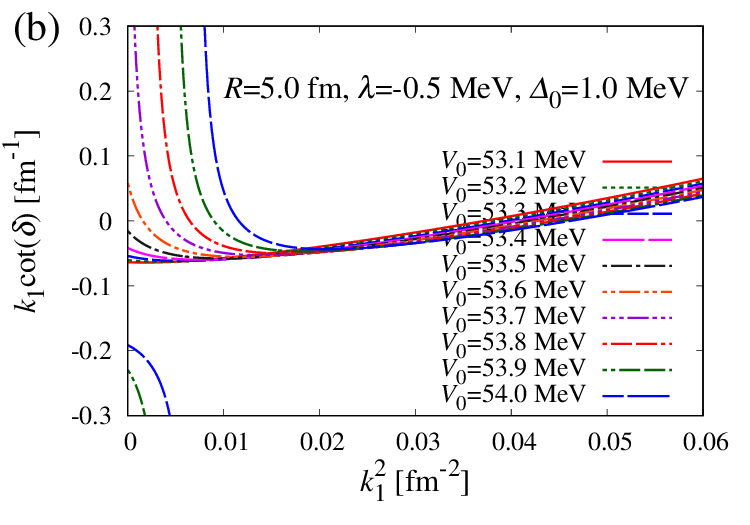}
\caption{The same as Fig.~8, but for $V_{0}=53.1-54.0$ MeV.}
\end{center}
\end{figure}

As shown in Fig.~10(b), the intercept of $k_{1}\cot\delta$, namely $-1/a$, changes from $-0$ to $+0$ in the range $V_{0}=53.5-53.6$ MeV, indicating the emergence of the hole-like unitary limit $U_{\mathrm{h}}$. This change originates from the divergence of $k_{1}\cot\delta$ caused by the zero crossing of $\delta$ associated with the quasiparticle resonance. The relationship between the quasiparticle resonance and the appearance of $U_{\mathrm{h}}$ will be discussed below in connection with the S-matrix pole trajectories.

Figure~10(b) also shows that the slope of $k_{1}\cot\delta$ at $U_{\mathrm{h}}$, corresponding to $r_{\mathrm{eff}}/2$, becomes negative. Table~3 lists the values of the potential depth $V_{0}$ at which $U_{\mathrm{h}}$ appears and the corresponding effective range $r_{\mathrm{eff}}$ for several values of the pairing strength $\Delta_{0}$. As seen in Table~3, the effective range at $U_{\mathrm{h}}$ takes large negative values.

\begin{table}[ht]
\caption{The same as Table.~2, but for the hole-like unitary limits $U_{\mathrm{h}}$.}
\begin{center}
\begin{tabular}{ccc} \hline
$\Delta_{0}$ [MeV] & $V_{0}$ [MeV] & $r_{\mathrm{eff}}$ [$\mathrm{fm}$]\\ \hline
0.2 & 53.936 & $-1788.903$\\
0.4 & 53.893 & $-429.550$\\
0.6 & 53.816 & $-177.571$\\
0.8 & 53.701 & $-89.029$\\
1.0 & 53.532 & $-47.527$\\ \hline
\end{tabular}
\end{center}
\end{table}

Figure~11 presents the behavior of the S-matrix pole around $V_{0}\sim 53.5$ MeV, where $U_{\mathrm{h}}$ appears. For $V_{0}\gtrsim 53.5$ MeV, the pole is located in the fourth quadrant of the $E^{(2)}$-sheet and gives rise to a hole-like quasiparticle resonance with a finite width. As $V_{0}$ decreases, the pole crosses the line $\mathrm{Re}(E)=-\lambda$ from right to left at $V_{0}=53.554$ MeV and subsequently passes through the threshold $E=(-\lambda,~0)$ at $V_{0}=53.532$ MeV, moving to the $E^{(1)}$-sheet. The hole-like unitary limit $U_{\mathrm{h}}$ emerges at this transition. Thus, the appearance of $U_{\mathrm{h}}$ is closely related to the presence of a hole-like quasiparticle resonance.

\begin{figure}[ht]
\begin{center}
\includegraphics[width=80mm]{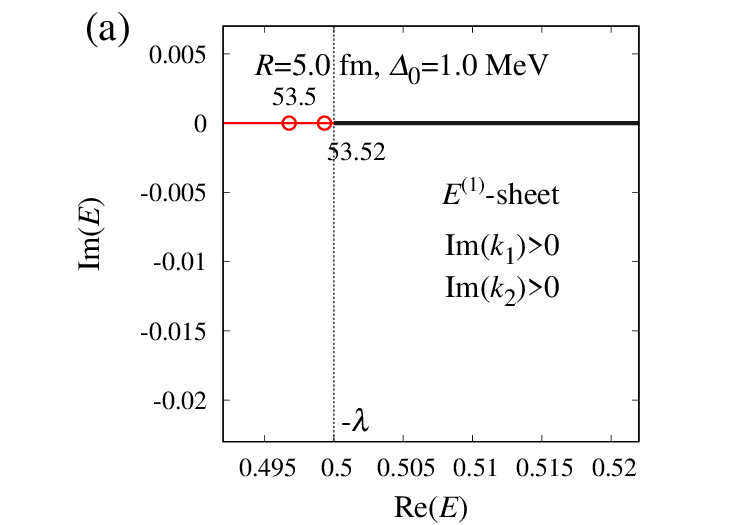}
\includegraphics[width=80mm]{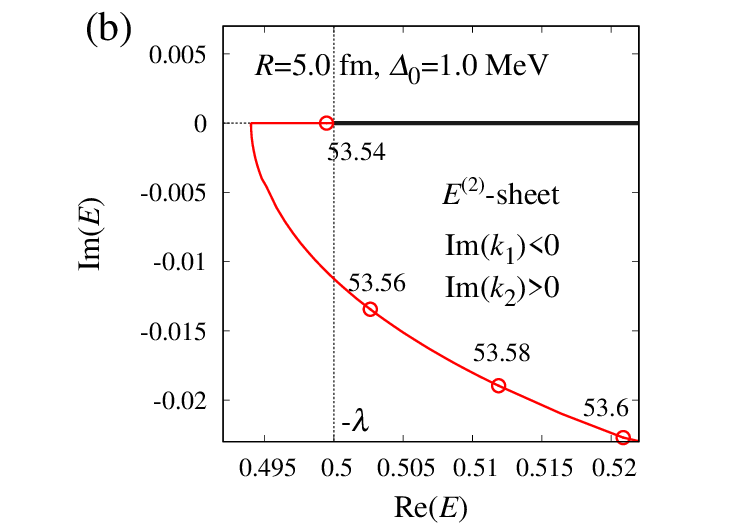}
\caption{The same as Fig.~9, but for $V_{0}=53.1-54.0$ MeV. The symbols are for every 0.02 MeV interval in $V_{0}$. Note that values near the convergence point were obtained through extrapolation calculation.}
\end{center}
\end{figure}

\subsection{The low-energy effective range expansion in the vicinity of the unitary limits $U_{\mathrm{p}}$ and $U_{\mathrm{h}}$}
\subsubsection{Weak pairing effects at $U_{\mathrm{p}}$}
As discussed above, the behavior of the phase shift and the S-matrix near the particle-like unitary limit $U_{\mathrm{p}}$ is qualitatively similar to that observed in single-particle elastic scattering. For example, the low-energy behavior of $\delta$ and $k_{1}\cot\delta$ near $U_{\mathrm{p}}$ shown in Fig.~8 closely resembles that of the single-particle scattering problem shown in Fig.~3. In addition, the dependence of $k_{1}\cot\delta$ on $k_{1}^{2}$ is approximately linear, suggesting that the effective range expansion provides a reasonable description in this regime.

Using the phase shift derived from the effective range expansion,
\begin{equation}
\delta\left( k_{1} \right)=\arctan\left[ k_{1}\left( -\frac{1}{a}+\frac{1}{2}r_{\mathrm{eff}}k^{2}_{1} \right)^{-1} \right],
\end{equation}
we substitute the scattering length $a$ and effective range $r_{\mathrm{eff}}$ obtained from Eq.~(12). The resulting phase shifts are shown in Fig.~12. As seen in the figure, for $e\lesssim 0.1$ MeV, the phase shifts obtained from Eq.~(24), denoted as `ERE', agree well with the results shown in Fig.~8(a) denoted as `Excat'. Even outside this very low-energy region, Eq.~(24) reproduces the overall trend of the phase shift reasonably well. These results indicate that, in the vicinity of $U_{\mathrm{p}}$, analyses based on the effective range expansion are valid.

\begin{figure}[ht]
\begin{center}
\includegraphics[width=80mm]{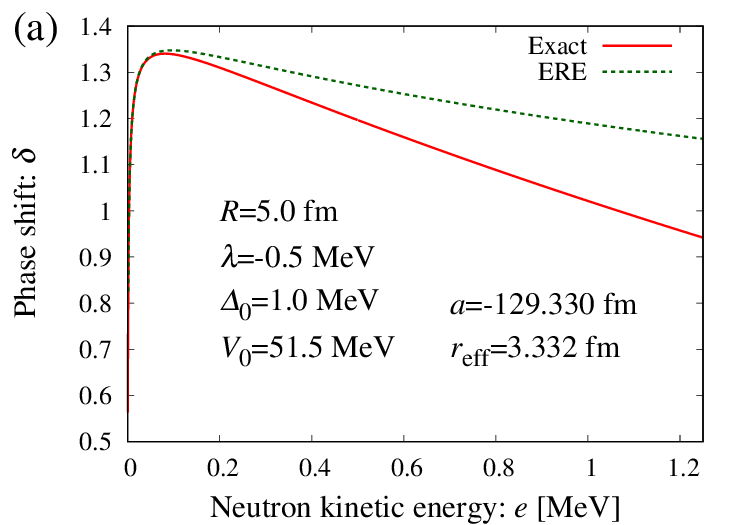}
\includegraphics[width=80mm]{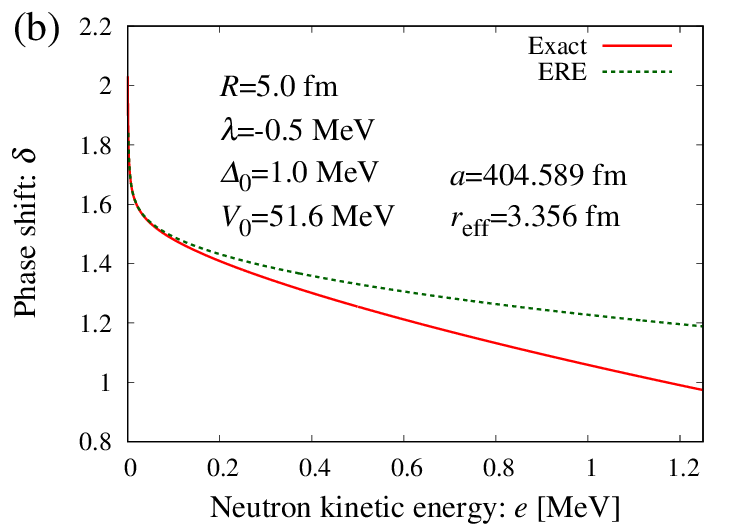}
\caption{`Excat' corresponds to the results shown in Fig.~8(a). `ERE' are the phase shifts from Eq.~(24) with calculated $a$ and $r_{\mathrm{eff}}$ for (a) $V_{0}=51.5$ MeV and (b) $V_{0}=51.6$ MeV. }
\end{center}
\end{figure}

Although pairing correlations reduce the effective range, as shown in Table~2, the overall magnitude of the pairing effects at $U_{\mathrm{p}}$ remains modest. Therefore, the particle-like unitary limit $U_{\mathrm{p}}$ can be regarded as a natural generalization of the unitary limit in single-particle elastic scattering.

\subsubsection{Strong pairing effects leading to $U_{\mathrm{h}}$}
The hole-like unitary limit $U_{\mathrm{h}}$ is a pairing-induced unitary limit associated with a hole-like quasiparticle resonance. As shown by the $V_{0}$ dependence of the S-matrix poles, $U_{\mathrm{h}}$ emerges when a hole-like quasiparticle resonance pole on the $E^{(2)}$-sheet crosses the threshold at $E=(-\lambda,0)$.

In the vicinity of $U_{\mathrm{h}}$, the slope of $k_{1}\cot\delta$ in the low-energy region becomes negative, corresponding to a negative effective range $r_{\mathrm{eff}}$ (Fig.~10(b)). Moreover, the dependence of $k_{1}\cot\delta$ on $k_{1}^{2}$ is no longer linear, indicating the breakdown of the low-energy effective range expansion. As summarized in Table~3, the effective range at $U_{\mathrm{h}}$ also shows a strong dependence on the pairing strength $\Delta_{0}$.

Figure~13 compares the phase shifts obtained from the effective range expansion in Eq.~(24), using the scattering length $a$ and effective range $r_{\mathrm{eff}}$ from Eq.~(12), with the analytically calculated results shown in Fig.~10(a). Except for an extremely narrow region near $e\sim 0$, the two results differ significantly. This clearly demonstrates that analyses based solely on the scattering length and effective range are not appropriate in the vicinity of $U_{\mathrm{h}}$.

\begin{figure}[ht]
\begin{center}
\includegraphics[width=80mm]{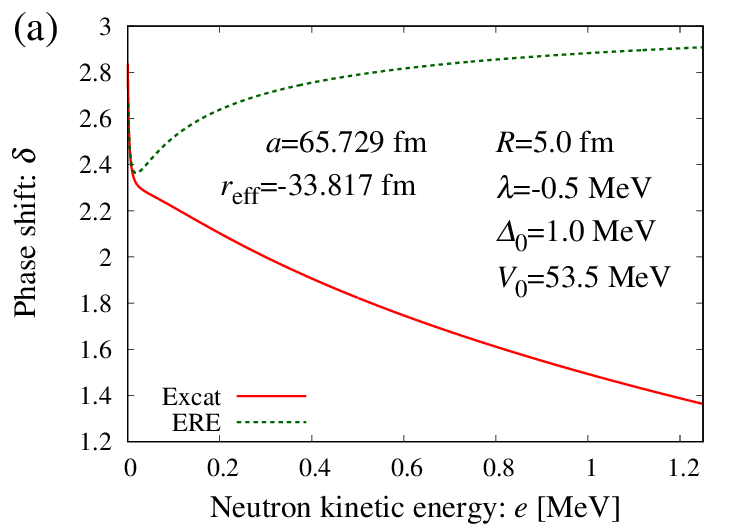}
\includegraphics[width=80mm]{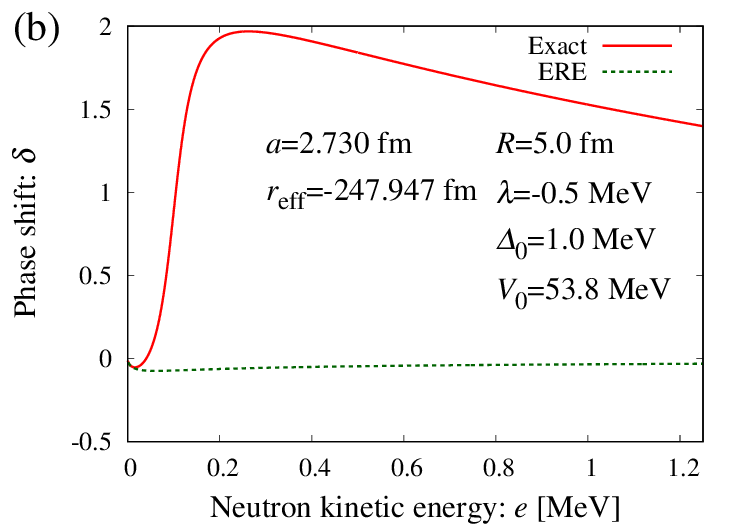}
\caption{The same as Fig.~12, but for (a) $V_{0}=53.5$ MeV and (b) $V_{0}=53.8$ MeV.}
\end{center}
\end{figure}

Given that $U_{\mathrm{h}}$ originates from hole-like quasiparticle resonances, a description explicitly incorporating resonance poles is expected to be more suitable. Following our previous study~\cite{Kobayashi2020}, we describe the phase shift using an approximate expression that incorporates contributions from four S-matrix poles, as given below.
\begin{equation}
S(k_{1})\sim S^{\text{pole}}(k_{1})=\frac{k_{1}-\bar{k}^{\ast}_{1a}}{k_{1}-\bar{k}_{1a}}\cdot\frac{k_{1}- \bar{k}^{\ast}_{1b}}{k_{1}-\bar{k}_{1b}}\cdot\frac{k_{1}-\bar{k}^{\ast}_{1\bar{a}}}{k_{1}-\bar{k}_{1\bar{a}}}\cdot\frac{k_{1}-\bar{k}^{\ast}_{1\bar{b}}}{k_{1}-\bar{k}_{1\bar{b}}}~,
\end{equation}
\begin{equation}
\delta^{\mathrm{pole}}(k_{1})=\mathrm{Re}\left[ \frac{1}{2i}\ln S^{\text{pole}}(k_{1}) \right].
\end{equation}

Table~4 lists the pole positions obtained from Eq.~(14) for $V_{0}=53.5$ MeV and $V_{0}=53.8$ MeV. The case $V_{0}=53.5$ MeV corresponds to a configuration in which all four poles lie on the imaginary axis of the $k_{1}$-plane, whereas $V_{0}=53.8$ MeV represents a resonance-like case with $\bar{k}_{1a}$ located in the fourth quadrant. In the latter case, the Schwarz reflection principle implies $\bar{k}_{1b}=-\bar{k}^{\ast}_{1a}$ and $\bar{k}_{1\bar{b}}=-\bar{k}^{\ast}_{1\bar{a}}$.

\begin{table}[ht]
\caption{Positions of the four poles $\bar{k}_{1a}$, $\bar{k}_{1b}$, $\bar{k}_{1\bar{a}}$, and $\bar{k}_{1\bar{b}}$ for $V_{0}=53.5$ MeV and $V_{0}=53.8$ MeV.}
\begin{center}
\begin{tabular}{ccccccc} \hline
$V_{0}$ [MeV] &~~~& $\bar{k}_{1a}$ [fm${}^{-1}$] & $\bar{k}_{1b}$ [fm${}^{-1}$] &~~~& $\bar{k}_{1\bar{a}}$ [fm${}^{-1}$] & $\bar{k}_{1\bar{b}}$ [fm${}^{-1}$] \\ \hline
53.5 & & $(0,~0.012)$ & $(0,~-0.045)$ & & $(0,~0.219)$ & $(0,~0.215)$\\
53.8 & & $(0.074,~-0.012)$ & $(-0.074,~-0.012)$ & & $(0.004,~0.232)$ & $(-0.004,~0.232)$\\ \hline
\end{tabular}
\end{center}
\end{table}

Figure 14 shows the positions of the four poles listed in Table~4 on the complex $E$-plane. For $V_{0}=53.5$ MeV (Fig.~14(a)), all four poles lie on the real axis. Here, poles $a$ and $\bar{a}$ are located on the $E^{(1)}$-sheet, pole $b$ on the $E^{(2)}$-sheet, and pole $\bar{b}$ on the $E^{(3)}$-sheet. On the other hand, for $V_{0}=53.8$ MeV (Fig.~14(b)), all four poles move away from the real axis and acquire a finite imaginary part. The poles $a$ and $b$ are located on the $E^{(2)}$-sheet while $\bar{a}$ and $\bar{b}$ are on the $E^{(3)}$-sheet.

\begin{figure}[ht]
\begin{center}
\includegraphics[width=80mm]{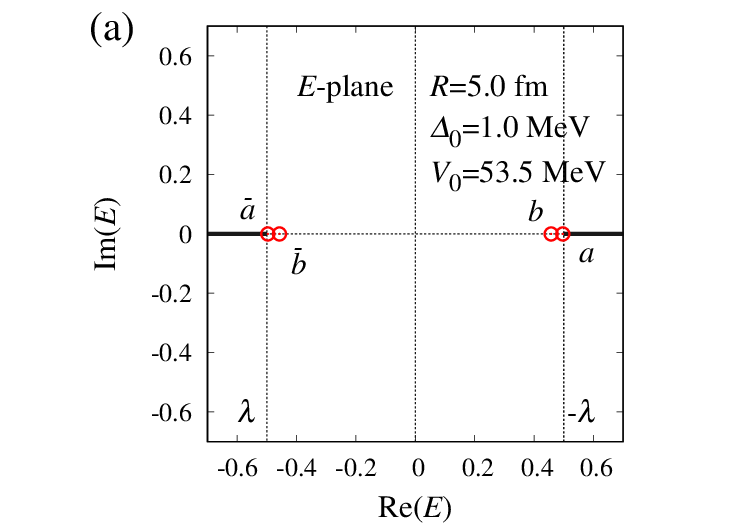}
\includegraphics[width=80mm]{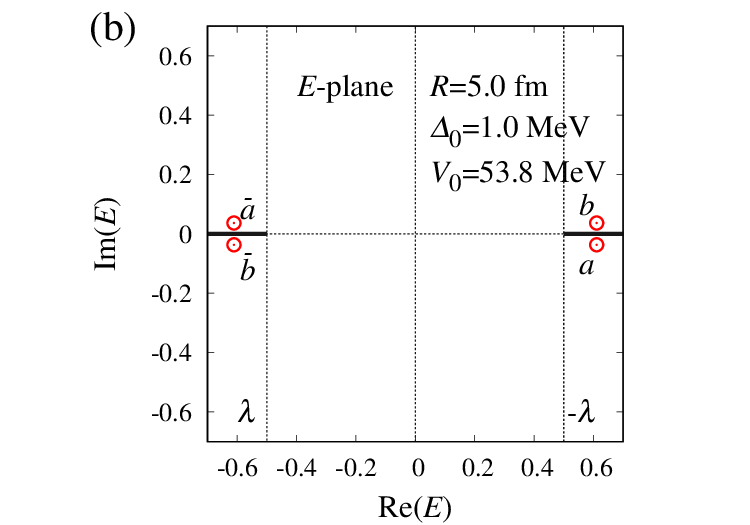}
\caption{Positions of the four poles for (a) $V_{0}=53.5$ MeV and (b) $V_{0}=53.8$ MeV in the complex $E$-plane. The two thick lines on the real $E$-axis are the branch cuts.}
\end{center}
\end{figure}

Figure~15 compares the phase shifts calculated from Eq.~(26) with the results shown in Fig.~10(a). When only the quasiparticle poles $\bar{k}_{1a}$ and $\bar{k}_{1b}$ are included, the results cannot be reproduced. In contrast, when the contributions from all four poles, including the quasihole poles $\bar{k}_{1\bar{a}}$ and $\bar{k}_{1\bar{b}}$, are taken into account, the behavior in Fig.~10(a) is well reproduced. This indicates that the quasihole poles contribute predominantly as a background component.

\begin{figure}[ht]
\begin{center}
\includegraphics[width=80mm]{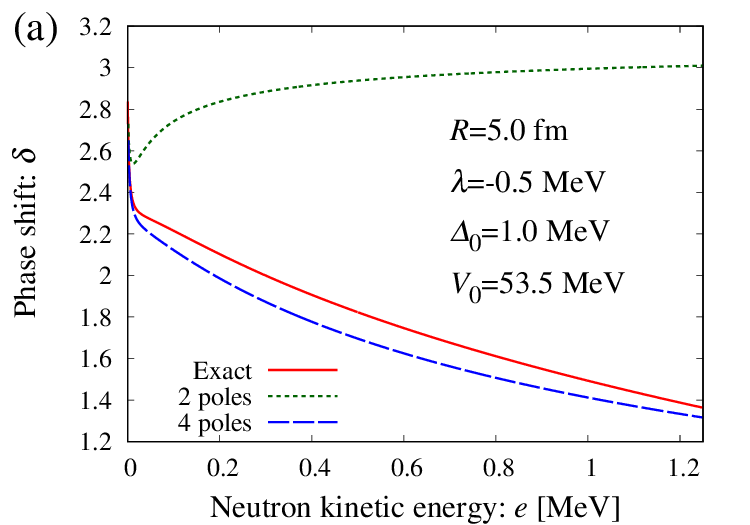}
\includegraphics[width=80mm]{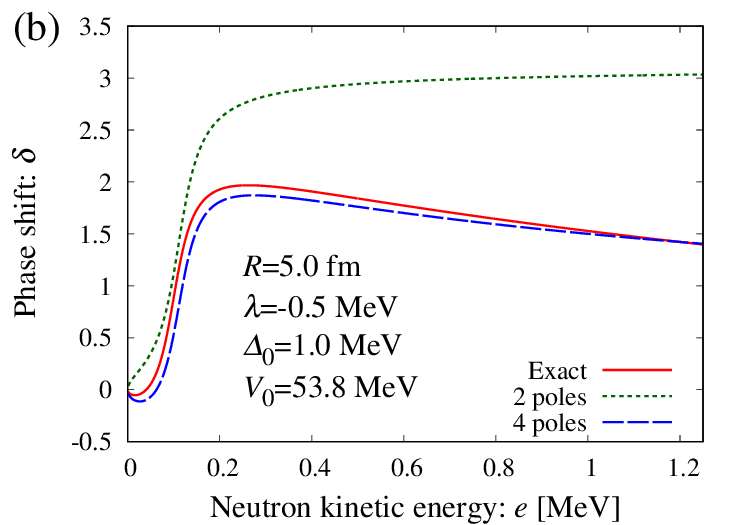}
\caption{`Excat' corresponds to the results shown in Fig.~10(a). `2 poles' are the phase shifts from Eq.~(26) with calculated $\bar{k}_{1a}$ and $\bar{k}_{1b}$. `4 poles' are the phase shifts from Eq.~(26) with calculated the four poles. (a) and (b) are the results for $V_{0}=53.5$ MeV and $V_{0}=53.8$ MeV, respectively. }
\end{center}
\end{figure}

In summary, pairing correlations induce a qualitative change by generating a new unitary limit $U_{\mathrm{h}}$ through hole components. In the presence of such a hole-like quasiparticle resonance-induced unitary limit, analyses based on the four poles description in Eqs.~(25) and~(26) are more appropriate than those relying on the low-energy effective range expansion.

\section{Conclusions}
In this study, we have examined in a comprehensive and systematic way influence of pairing effects on low-energy $s$-wave quasiparticle elastic scattering by employing a schematic square-well model within the coordinate space Hartree-Fock-Bogoliubov theory. Analytical expressions for the phase shift, elastic cross section, scattering length, effective range, and S-matrix were derived, allowing for a systematic investigation of pairing effects.

A central result of this work is the demonstration that pairing correlations give rise to two distinct unitary limits. The particle-like unitary limit $U_{\mathrm{p}}$ represents a natural extension of the conventional unitary limit of single-particle scattering, for which the effective range expansion remains valid. In contrast, the hole-like unitary limit $U_{\mathrm{h}}$ emerges as a pairing-induced phenomenon associated with quasiparticle resonances and exhibits qualitatively different scattering behavior.

In the vicinity of the particle-like unitary limit $U_{\mathrm{p}}$, the behavior of the phase shift and the S-matrix poles remains qualitatively similar to that of the single-particle scattering problem. Although the effective range $r_{\mathrm{eff}}$ decreases monotonically with increasing pairing strength $\Delta_{0}$, the overall pairing effects are modest, and the low-energy effective range expansion provides a reliable description. From this viewpoint, $U_{\mathrm{p}}$ can be regarded as a natural extension of the single-particle unitary limit to superfluid systems.

In contrast, the hole-like unitary limit $U_{\mathrm{h}}$ exhibits qualitatively different scattering characteristics. In this regime, the effective range takes large negative values and the low-energy quasiparticle scattering shows a drastic energy dependence. Consequently, the effective range expansion breaks down and fails to capture the essential physics near $U_{\mathrm{h}}$. Instead, we showed that an approximate description explicitly incorporating contributions from four S-matrix poles associated with quasiparticle resonances provides an appropriate and physically transparent description. This clearly illustrates that, under strong pairing correlations, pole-based approaches are indispensable for understanding low-energy $s$-wave quasiparticle scattering.

By adopting a shallow Fermi energy, the present analysis was designed to mimic weakly bound nuclei near the drip-line. It is well known that exotic phenomena such as halo formation and Efimov states can be enhanced in the vicinity of the unitary limit in such systems. The particle-like and hole-like unitary limits identified in this work may therefore be relevant to these phenomena. An important next step is to examine to what extent the qualitative features revealed by this schematic model persist in realistic drip-line nuclei.

Finally, we note that the present study is based on a simplified model employing square-well Hartree-Fock and pairing potentials with a common radius, and does not include important effects such as spin-orbit coupling or nuclear deformation. As a result, direct quantitative comparison with experimental data is beyond the scope of this work. Future studies using more realistic descriptions, such as Woods-Saxon potentials or fully self-consistent Hartree-Fock-Bogoliubov calculations, are highly desirable. Extensions to reaction processes, including neutron capture and breakup reactions, where unitary-limit behavior may play a significant role, also represent promising directions for future research.

\section*{Acknowledgment}
This work is supported by the JSPS KAKENHI, Grant No. JP23K13104 and No. JP24K07014.

\appendix

\section{The expression of $k_{\pm}$ and $\beta$ with $E<\Delta_{0}$}
For quasiparticle energies below the pairing gap, $E<\Delta_{0}$, the wave numbers $k_{\pm}$ and the coefficient $\beta$ become complex. To distinguish this case from the region $E\ge\Delta_{0}$, we denote the corresponding quantities for $E<\Delta_{0}$ by primes. They are given by
\begin{equation}
k^{\prime}_{\pm}=\sqrt{ \frac{2m}{\hbar^{2}} \left(V_{0}+\lambda \pm i\sqrt{\Delta^{2}_{0}-E^{2}} \right) },\quad \beta^{\prime}=\frac{\Delta_{0}}{E+i\sqrt{\Delta^{2}_{0}-E^{2}}}\quad (E<\Delta_{0}).
\end{equation}

From the above definitions, one finds the relation $k^{\prime\ast}_{+}=k^{\prime}_{-}$, which leads to
\begin{equation}
\mathrm{Re}\left( k^{\prime}_{+} \right)=\mathrm{Re}\left( k^{\prime}_{-} \right),\quad \mathrm{Im}\left( k^{\prime}_{+} \right)=-\mathrm{Im}\left( k^{\prime}_{-} \right)~.
\end{equation}

The real and imaginary parts of $k^{\prime}_{+}$ are explicitly given by
\begin{equation}
\mathrm{Re}\left( k^{\prime}_{+} \right)=\sqrt{\frac{m}{\hbar^{2}}}\sqrt{(V_{0}+\lambda)+\sqrt{(V_{0}+\lambda)^{2}+\left( \Delta^{2}_{0}-E^{2} \right)}},\quad \mathrm{Im}\left( k^{\prime}_{+} \right)=\frac{m}{\hbar^2}\sqrt{\Delta^{2}_{0}-E^{2}}\left[ \mathrm{Re}\left( k^{\prime}_{+} \right) \right]^{-1}~.
\end{equation}

In the limit $E\to\Delta_{0}$, these quantities satisfy
\begin{equation}
\lim_{E\to\Delta_{0}}\mathrm{Re}\left( k^{\prime}_{+} \right)=\sqrt{\frac{2m}{\hbar^2}(V_{0}+\lambda)}=\lim_{E\to\Delta_{0}}k_{\pm},\quad \lim_{E\to\Delta_{0}}\mathrm{Im}\left( k^{\prime}_{+} \right)=0~.
\end{equation}

The relations between $k_{1,2}$ and $k^{\prime}_{\pm}$ are analogous to those for $k_{\pm}$, namely,
\begin{equation}
\lim_{V_{0},\Delta_{0}\to0}k_{+}=\lim_{V_{0},\Delta_{0}\to0}k^{\prime}_{+}=k_{1},\quad \lim_{V_{0},\Delta_{0}\to0}k_{-}=\lim_{V_{0},\Delta_{0}\to0}k^{\prime}_{-}=k_{2}~.
\end{equation}

The real and imaginary parts of $\beta^{\prime}$ are obtained as
\begin{equation}
\mathrm{Re}(\beta^{\prime})=\frac{E}{\Delta_{0}},\quad \mathrm{Im}(\beta^{\prime})=-\frac{\sqrt{\Delta^{2}_{0}-E^{2}}}{\Delta_{0}}~.
\end{equation}

In the limit $E\to\Delta_{0}$, one recovers
\begin{equation}
\lim_{E\to\Delta_{0}}\mathrm{Re}(\beta^{\prime})=\lim_{E\to\Delta_{0}}\beta=1,\quad \lim_{E\to\Delta_{0}}\mathrm{Im}(\beta^{\prime})=0
\end{equation}
which is consistent with the expressions for $E\ge\Delta_{0}$.

For $E<\Delta_{0}$, the quasiparticle wave function inside the well can be written as
\begin{eqnarray}
\left(
\begin{array}{c}
u^{\prime}_{\mathrm{in}}(r) \\
v^{\prime}_{\mathrm{in}}(r)
\end{array}
\right)&=&A\left(
\begin{array}{c}
1 \\
\beta^{\prime}
\end{array}
\right)\sin k^{\prime}_{+}r+B\left(
\begin{array}{c}
\beta^{\prime} \\
1
\end{array}
\right)\sin k^{\prime}_{-}r\\
&=&\left(
\begin{array}{c}
A^{\prime} \\
\mathrm{Re}(\beta^{\prime})A^{\prime}+\mathrm{Im}(\beta^{\prime})B^{\prime}
\end{array}
\right)\cosh \mathrm{Im}(k^{\prime}_{+})r\sin \mathrm{Re}(k^{\prime}_{+})r\nonumber\\
&&+\left(
\begin{array}{c}
B^{\prime} \\
-\mathrm{Im}(\beta^{\prime})A^{\prime}+\mathrm{Re}(\beta^{\prime})B^{\prime}
\end{array}
\right)\sinh \mathrm{Im}(k^{\prime}_{+})r\cos \mathrm{Re}(k^{\prime}_{+})r~.
\end{eqnarray}
Here, the primes attached to $u^{\prime}_{\mathrm{in}}(r)$ and $v^{\prime}_{\mathrm{in}}(r)$ also indicate the case $E<\Delta_{0}$ and do not represent derivatives.

\section{The expression of $\mathcal{K}$}
For $E\ge\Delta_{0}$, the quantity $\mathcal{K}$ is given by
\begin{equation}
\mathcal{K}=\frac{k_{-}\cos k_{-}R\cdot X-k_{+}\cos k_{+}R\cdot Y}{\sin k_{-}R\cdot X-\sin k_{+}R\cdot Y}~.
\end{equation}
where
\begin{equation}
X=\beta^{2}\left( k_{+}\cos k_{+}R+\kappa_{2}\sin k_{+}R \right)~,
\end{equation}
\begin{equation}
Y=k_{-}\cos k_{-}R+\kappa_{2}\sin k_{-}R~.
\end{equation}

In the limit $\Delta_{0}\to 0$, one recovers
\begin{equation}
\lim_{\Delta_{0}\to0}\mathcal{K}=k_{\mathrm{in}}\cot k_{\mathrm{in}}R~.
\end{equation}

For $E<\Delta_{0}$, the corresponding quantity $\mathcal{K}^{\prime}$ is defined as
\begin{equation}
\mathcal{K}^{\prime}=\frac{\nu X^{\prime}+\mu Y^{\prime}}{\zeta X^{\prime}+\eta Y^{\prime}}
\end{equation}
where the coefficients $\zeta$, $\eta$, $\mu$, $\nu$, $X^{\prime}$, and $Y^{\prime}$ are given explicitly in Eqs.~(B6)–(B11).
\begin{equation}
\zeta=\cosh\left[ \mathrm{Im}\left( k^{\prime}_{+} \right)R \right]\sin\left[ \mathrm{Re}\left( k^{\prime}_{+} \right)R \right]
\end{equation}
\begin{equation}
\eta=\sinh\left[ \mathrm{Im}\left( k^{\prime}_{+} \right)R \right]\cos\left[ \mathrm{Re}\left( k^{\prime}_{+} \right)R \right]
\end{equation}
\begin{equation}
\mu=\mathrm{Im}\left( k^{\prime}_{+} \right)\cosh\left[ \mathrm{Im}\left( k^{\prime}_{+} \right)R \right]\cos\left[ \mathrm{Re}\left( k^{\prime}_{+} \right)R \right]-\mathrm{Re}\left( k^{\prime}_{+} \right)\sinh\left[ \mathrm{Im}\left( k^{\prime}_{+} \right)R \right]\sin\left[ \mathrm{Re}\left( k^{\prime}_{+} \right)R \right]
\end{equation}
\begin{equation}
\nu=\mathrm{Im}\left( k^{\prime}_{+} \right)\sinh\left[ \mathrm{Im}\left( k^{\prime}_{+} \right)R \right]\sin\left[ \mathrm{Re}\left( k^{\prime}_{+} \right)R \right]+\mathrm{Re}\left( k^{\prime}_{+} \right)\cosh\left[ \mathrm{Im}\left( k^{\prime}_{+} \right)R \right]\cos\left[ \mathrm{Re}\left( k^{\prime}_{+} \right)R \right]
\end{equation}
\begin{equation}
X^{\prime}=\mathrm{Re}(\beta^{\prime})\mu+\mathrm{Im}(\beta^{\prime})\nu+\kappa_{2}\left[ \mathrm{Re}(\beta^{\prime})\eta+\mathrm{Im}(\beta^{\prime})\zeta \right]
\end{equation}
\begin{equation}
Y^{\prime}=\mathrm{Im}(\beta^{\prime})\mu-\mathrm{Re}(\beta^{\prime})\nu-\kappa_{2}\left[ -\mathrm{Im}(\beta^{\prime})\eta+\mathrm{Re}(\beta^{\prime})\zeta \right]
\end{equation}

Furthermore, in the limit $E\to\Delta_{0}$, one finds
\begin{eqnarray}
\lim_{E\to\Delta_{0}}\mathcal{K}=\lim_{E\to\Delta_{0}}\mathcal{K}^{\prime}=0~.
\end{eqnarray}

As a consequence, Eq.~(10) yields
\begin{equation}
\lim_{E\to\Delta_{0}}\tan\delta=\lim_{E\to\Delta_{0}}\tan\delta^{\prime}=\cot k_{1}R~.
\end{equation}

\section{The expression of $\tilde{\mathcal{K}}$ and $\mathcal{R}$}
To derive $\tilde{\mathcal{K}}$ and $\mathcal{R}$, we expand $\mathcal{K}$ with respect to $k_{1}^{2}$ as
\begin{equation}
\mathcal{K}=\tilde{\mathcal{K}}+\mathcal{R}k^{2}_{1}~.
\end{equation}

As in the main text, all quantities with a tilde stand for the $k_{1}\to0$ limits, i.e. the constant terms of the expansions in $k^{2}_{1}$, while $A,~B,~C,\ldots$ denote the corresponding coefficients of $k^{2}_{1}$.For composite quantities, the tilde is placed over the whole symbol, as in $\widetilde{\mathrm{Re}\left( k^{\prime}_{+} \right)}$

For $E\ge\Delta_{0}$, this expansion is obtained by expanding $k_{\pm}$, $\kappa_{2}$, $\beta$, $X$, and $Y$ up to second order in $k_{1}^{2}$, leading to explicit expressions for $\tilde{\mathcal{K}}$ and $\mathcal{R}$ given in Eqs.~(C8) and~(C9).

\begin{equation}
\kappa_{2}=\tilde{\kappa}_{2}+Ak^{2}_{1},\quad \tilde{\kappa}_{2}=\sqrt{-\frac{4m\lambda}{\hbar^2}},\quad A=\frac{1}{2\tilde{\kappa}_{2}}
\end{equation}
\begin{eqnarray}
&&k_{+}=\tilde{k}_{+}+Bk^{2}_{1},\quad \tilde{k}_{+}=\sqrt{\frac{2m}{\hbar^2}\left( V_{0}+\lambda+\sqrt{\lambda^{2}-\Delta^{2}_{0}} \right)},\nonumber\\
&&B=-\frac{1}{4\sqrt{\left( \lambda^{2}-\Delta^{2}_{0} \right)\left( V_{0}+\lambda+\sqrt{\lambda^{2}-\Delta^{2}_{0}} \right)}}\sqrt{\frac{2m}{\hbar^2}}\frac{\hbar^{2}\lambda}{m}
\end{eqnarray}
\begin{eqnarray}
&&k_{-}=\tilde{k}_{-}+Ck^{2}_{1},\quad \tilde{k}_{-}=\sqrt{\frac{2m}{\hbar^2}\left( V_{0}+\lambda-\sqrt{\lambda^{2}-\Delta^{2}_{0}} \right)},\nonumber\\
&&C=\frac{1}{4\sqrt{\left( \lambda^{2}-\Delta^{2}_{0} \right)\left( V_{0}+\lambda-\sqrt{\lambda^{2}-\Delta^{2}_{0}} \right)}}\sqrt{\frac{2m}{\hbar^2}}\frac{\hbar^{2}\lambda}{m}
\end{eqnarray}
\begin{eqnarray}
&&\beta=\tilde{\beta}+Dk^{2}_{1},\quad \tilde{\beta}=\frac{\Delta_{0}}{-\lambda+\sqrt{\lambda^{2}-\Delta^{2}_{0}}},\nonumber\\
&&D=-\frac{\Delta_{0}}{\left( -\lambda+\sqrt{\lambda^{2}-\Delta^{2}_{0}} \right)^2}\left( \frac{\hbar^2}{2m}-\frac{1}{2\sqrt{\lambda^{2}-\Delta^{2}_{0}}}\frac{\hbar^{2}\lambda}{m} \right)
\end{eqnarray}
\begin{eqnarray}
&&X=\tilde{X}+Fk^{2}_{1},\nonumber\\
&&\tilde{X}=\tilde{\beta}^{2}\left( \tilde{k}_{+}\cos \tilde{k}_{+}R+\tilde{\kappa}_{2}\sin \tilde{k}_{+}R \right),\nonumber\\
&&F=\tilde{\beta}^{2}\left( -B\tilde{k}_{+}R\sin \tilde{k}_{+}R+B\cos \tilde{k}_{+}R+B\tilde{\kappa}_{2}R\cos \tilde{k}_{+}R+A\sin \tilde{k}_{+}R \right)\nonumber\\
&&\quad\quad+2\tilde{\beta}D\left( \tilde{k}_{+}\cos \tilde{k}_{+}R+\tilde{\kappa}_{2}\sin \tilde{k}_{+}R \right)
\end{eqnarray}
\begin{eqnarray}
&&Y=\tilde{Y}+Gk^{2}_{1},\nonumber\\
&&\tilde{Y}=\tilde{k}_{-}\cos \tilde{k}_{-}R+\tilde{\kappa}_{2}\sin \tilde{k}_{-}R,\nonumber\\
&&G=-C\tilde{k}_{-}R\sin \tilde{k}_{-}R+C\cos \tilde{k}_{-}R+C\tilde{\kappa}_{2}R\cos \tilde{k}_{-}R+A\sin \tilde{k}_{-}R
\end{eqnarray}

\begin{equation}
\tilde{\mathcal{K}}=\frac{\tilde{k}_{-}\cos \tilde{k}_{-}R\cdot\tilde{X}-\tilde{k}_{+}\cos \tilde{k}_{+}R\cdot\tilde{Y}}{\sin\tilde{k}_{-}R\cdot\tilde{X}-\sin\tilde{k}_{+}R\cdot\tilde{Y}}
\end{equation}
\begin{eqnarray}
\mathcal{R}&=&-\tilde{\mathcal{K}}\cdot\frac{CR\cos \tilde{k}_{-}R\cdot\tilde{X}+F\sin \tilde{k}_{-}R-BR\cos \tilde{k}_{+}R\cdot\tilde{Y}-G\sin\tilde{k}_{+}R}{\sin\tilde{k}_{-}R\cdot\tilde{X}-\sin\tilde{k}_{+}R\cdot\tilde{Y}}\nonumber\\
&&+\frac{1}{\sin\tilde{k}_{-}R\cdot\tilde{X}-\sin\tilde{k}_{+}R\cdot\tilde{Y}}\left( -C\tilde{k}_{-}R\sin\tilde{k}_{-}R\cdot\tilde{X}+C\cos\tilde{k}_{-}R\cdot\tilde{X}+F\tilde{k}_{-}\cos\tilde{k}_{-}R \right.\nonumber\\
&&\left. \quad+B\tilde{k}_{+}R\sin\tilde{k}_{+}R\cdot\tilde{Y}-B\cos\tilde{k}_{+}R\cdot\tilde{Y}-G\tilde{k}_{+}\cos\tilde{k}_{+}R \right)
\end{eqnarray}

For $E<\Delta_{0}$, a similar expansion is carried out using the real and imaginary parts of $k^{\prime}_{+}$ and $\beta^{\prime}$. The resulting expansion
\begin{equation}
\mathcal{K}^{\prime}=\tilde{\mathcal{K}}^{\prime}+\mathcal{R}^{\prime}k^{2}_{1}~,
\end{equation}
yields the coefficients $\tilde{\mathcal{K}}^{\prime}$ and $\mathcal{R}^{\prime}$, whose explicit forms are summarized in Eqs.~(C21) and~(C22).
\begin{eqnarray}
&&\mathrm{Re}\left( k^{\prime}_{+} \right)=\widetilde{\mathrm{Re}\left( k^{\prime}_{+} \right)}+B^{\prime}k^{2}_{1},\quad \widetilde{\mathrm{Re}\left( k^{\prime}_{+} \right)}=\sqrt{\frac{m}{\hbar^2}\left( V_{0}+\lambda+\sqrt{(V_{0}+\lambda)^{2}+\Delta^{2}_{0}-\lambda^{2}} \right)}\nonumber\\
&&B^{\prime}=\frac{\lambda}{ 4\widetilde{\mathrm{Re}\left( k^{\prime}_{+} \right)}\sqrt{(V_{0}+\lambda)^{2}+\Delta^{2}_{0}-\lambda^{2}} }
\end{eqnarray}
\begin{eqnarray}
&&\mathrm{Im}\left( k^{\prime}_{+} \right)=\widetilde{\mathrm{Im}\left( k^{\prime}_{+} \right)}+C^{\prime}k^{2}_{1},\quad \widetilde{\mathrm{Im}\left( k^{\prime}_{+} \right)}=\frac{m}{\hbar^2}\frac{\sqrt{\Delta^{2}_{0}-\lambda^{2}}}{\widetilde{\mathrm{Re}\left( k^{\prime}_{+} \right)}}\nonumber\\
&&C^{\prime}=-\frac{\widetilde{\mathrm{Im}\left( k^{\prime}_{+} \right)}}{\widetilde{\mathrm{Re}\left( k^{\prime}_{+} \right)}}B^{\prime}+\frac{\lambda}{2\widetilde{\mathrm{Re}\left( k^{\prime}_{+} \right)}\sqrt{\Delta^{2}_{0}-\lambda^{2}}}
\end{eqnarray}
\begin{equation}
\mathrm{Re}(\beta^{\prime})=\widetilde{\mathrm{Re}(\beta^{\prime})}+D^{\prime}k^{2}_{1},\quad \widetilde{\mathrm{Re}(\beta^{\prime})}=-\frac{\lambda}{\Delta_{0}},\quad D^{\prime}=\frac{\hbar^2}{2m\Delta_{0}}
\end{equation}
\begin{equation}
\mathrm{Im}(\beta^{\prime})=\widetilde{\mathrm{Im}(\beta^{\prime})}+F^{\prime}k^{2}_{1},\quad \widetilde{\mathrm{Im}(\beta^{\prime})}=-\frac{\sqrt{\Delta^{2}_{0}-\lambda^{2}}}{\Delta_{0}},\quad F^{\prime}=-\frac{\hbar^{2}\lambda}{2m\Delta_{0}\sqrt{\Delta^{2}_{0}-\lambda^{2}}}
\end{equation}
\begin{eqnarray}
&&\zeta=\tilde{\zeta}+G^{\prime}k^{2}_{1},\quad \tilde{\zeta}=\cosh\left[ \widetilde{\mathrm{Im}\left( k^{\prime}_{+} \right)}R \right]\sin\left[ \widetilde{\mathrm{Re}\left( k^{\prime}_{+} \right)}R \right],\nonumber\\
&&G^{\prime}=B^{\prime}R\cosh\left[ \widetilde{\mathrm{Im}\left( k^{\prime}_{+} \right)}R \right]\cos\left[ \widetilde{\mathrm{Re}\left( k^{\prime}_{+} \right)}R \right]+C^{\prime}R\sinh\left[ \widetilde{\mathrm{Im}\left( k^{\prime}_{+} \right)}R \right]\sin\left[ \widetilde{\mathrm{Re}\left( k^{\prime}_{+} \right)}R \right]
\end{eqnarray}
\begin{eqnarray}
&&\eta=\tilde{\eta}+H^{\prime}k^{2}_{1},\quad \tilde{\eta}=\sinh\left[ \widetilde{\mathrm{Im}\left( k^{\prime}_{+} \right)}R \right]\cos\left[ \widetilde{\mathrm{Re}\left( k^{\prime}_{+} \right)}R \right],\nonumber\\
&&H^{\prime}=-B^{\prime}R\sinh\left[ \widetilde{\mathrm{Im}\left( k^{\prime}_{+} \right)}R \right]\sin\left[ \widetilde{\mathrm{Re}\left( k^{\prime}_{+} \right)}R \right]+C^{\prime}R\cosh\left[ \widetilde{\mathrm{Im}\left( k^{\prime}_{+} \right)}R \right]\cos\left[ \widetilde{\mathrm{Re}\left( k^{\prime}_{+} \right)}R \right]
\end{eqnarray}
\begin{eqnarray}
&&\mu=\tilde{\mu}+I^{\prime}k^{2}_{1},\nonumber\\
&&\tilde{\mu}=\widetilde{\mathrm{Im}\left( k^{\prime}_{+} \right)}\cosh\left[ \widetilde{\mathrm{Im}\left( k^{\prime}_{+} \right)}R \right]\cos\left[ \widetilde{\mathrm{Re}\left( k^{\prime}_{+} \right)}R \right]-\widetilde{\mathrm{Re}\left( k^{\prime}_{+} \right)}\sinh\left[ \widetilde{\mathrm{Im}\left( k^{\prime}_{+} \right)}R \right]\sin\left[ \widetilde{\mathrm{Re}\left( k^{\prime}_{+} \right)}R \right]\nonumber\\
&&I^{\prime}=-B^{\prime}R\left( \widetilde{\mathrm{Im}\left( k^{\prime}_{+} \right)}\tilde{\zeta}+\widetilde{\mathrm{Re}\left( k^{\prime}_{+} \right)}\tilde{\eta}+\frac{1}{R}\sinh\left[ \widetilde{\mathrm{Im}\left( k^{\prime}_{+} \right)}R \right]\sin\left[ \widetilde{\mathrm{Re}\left( k^{\prime}_{+} \right)}R \right] \right)\nonumber\\
&&\quad\quad\quad+C^{\prime}R\left( \widetilde{\mathrm{Im}\left( k^{\prime}_{+} \right)}\tilde{\eta}-\widetilde{\mathrm{Re}\left( k^{\prime}_{+} \right)}\tilde{\zeta}+\frac{1}{R}\cosh\left[ \widetilde{\mathrm{Im}\left( k^{\prime}_{+} \right)}R \right]\cos\left[ \widetilde{\mathrm{Re}\left( k^{\prime}_{+} \right)}R \right] \right)
\end{eqnarray}
\begin{eqnarray}
&&\nu=\tilde{\nu}+J^{\prime}k^{2}_{1},\nonumber\\
&&\tilde{\nu}=\widetilde{\mathrm{Im}\left( k^{\prime}_{+} \right)}\sinh\left[ \widetilde{\mathrm{Im}\left( k^{\prime}_{+} \right)}R \right]\sin\left[ \widetilde{\mathrm{Re}\left( k^{\prime}_{+} \right)}R \right]+\widetilde{\mathrm{Re}\left( k^{\prime}_{+} \right)}\cosh\left[ \widetilde{\mathrm{Im}\left( k^{\prime}_{+} \right)}R \right]\cos\left[ \widetilde{\mathrm{Re}\left( k^{\prime}_{+} \right)}R \right]\nonumber\\
&&J^{\prime}=B^{\prime}R\left( \widetilde{\mathrm{Im}\left( k^{\prime}_{+} \right)}\tilde{\eta}-\widetilde{\mathrm{Re}\left( k^{\prime}_{+} \right)}\tilde{\zeta}+\frac{1}{R}\cosh\left[ \widetilde{\mathrm{Im}\left( k^{\prime}_{+} \right)}R \right]\cos\left[ \widetilde{\mathrm{Re}\left( k^{\prime}_{+} \right)}R \right] \right)\nonumber\\
&&\quad\quad\quad+C^{\prime}R\left( \widetilde{\mathrm{Im}\left( k^{\prime}_{+} \right)}\tilde{\zeta}+\widetilde{\mathrm{Re}\left( k^{\prime}_{+} \right)}\tilde{\eta}+\frac{1}{R}\sinh\left[ \widetilde{\mathrm{Im}\left( k^{\prime}_{+} \right)}R \right]\sin\left[ \widetilde{\mathrm{Re}\left( k^{\prime}_{+} \right)}R \right] \right)
\end{eqnarray}
\begin{eqnarray}
&&X^{\prime}=\tilde{X}^{\prime}+L^{\prime}k^{2}_{1},\quad \tilde{X}^{\prime}=\widetilde{\mathrm{Re}(\beta^{\prime})}\tilde{\mu}+\widetilde{\mathrm{Im}(\beta^{\prime})}\tilde{\nu}+\tilde{\kappa}_{2}\left[ \widetilde{\mathrm{Re}(\beta^{\prime})}\tilde{\eta}+\widetilde{\mathrm{Im}(\beta^{\prime})}\tilde{\zeta} \right]\nonumber\\
&&L^{\prime}=I^{\prime}\widetilde{\mathrm{Re}(\beta^{\prime})}+D^{\prime}\tilde{\mu}+J^{\prime}\widetilde{\mathrm{Im}(\beta^{\prime})}+F^{\prime}\tilde{\nu}+A\left( \widetilde{\mathrm{Re}(\beta^{\prime})}\tilde{\eta}+\widetilde{\mathrm{Im}(\beta^{\prime})}\tilde{\zeta} \right)\nonumber\\
&&\quad\quad+\tilde{\kappa}_{2}\left( H^{\prime}\widetilde{\mathrm{Re}(\beta^{\prime})}+D^{\prime}\tilde{\eta}+G^{\prime}\widetilde{\mathrm{Im}(\beta^{\prime})}+F^{\prime}\tilde{\zeta} \right)
\end{eqnarray}
\begin{eqnarray}
&&Y^{\prime}=\tilde{Y}^{\prime}+M^{\prime}k^{2}_{1},\quad \tilde{Y}^{\prime}=\widetilde{\mathrm{Im}(\beta^{\prime})}\tilde{\mu}-\widetilde{\mathrm{Re}(\beta^{\prime})}\tilde{\nu}-\tilde{\kappa}_{2}\left[ -\widetilde{\mathrm{Im}(\beta^{\prime})}\tilde{\eta}+\widetilde{\mathrm{Re}(\beta^{\prime})}\tilde{\zeta} \right]\nonumber\\
&&M^{\prime}=I^{\prime}\widetilde{\mathrm{Im}(\beta^{\prime})}+F^{\prime}\tilde{\mu}-J^{\prime}\widetilde{\mathrm{Re}(\beta^{\prime})}-D^{\prime}\tilde{\nu}-A\left( -\widetilde{\mathrm{Im}(\beta^{\prime})}\tilde{\eta}+\widetilde{\mathrm{Re}(\beta^{\prime})}\tilde{\zeta} \right)\nonumber\\
&&\quad\quad-\tilde{\kappa}_{2}\left( -H^{\prime}\widetilde{\mathrm{Im}(\beta^{\prime})}-F^{\prime}\tilde{\eta}+G^{\prime}\widetilde{\mathrm{Re}(\beta^{\prime})}+D^{\prime}\tilde{\zeta} \right)
\end{eqnarray}

\begin{equation}
\tilde{\mathcal{K}}^{\prime}=\frac{\tilde{\nu}\tilde{X}^{\prime}+\tilde{\mu}\tilde{Y}^{\prime}}{\tilde{\zeta}\tilde{X}^{\prime}+\tilde{\eta}\tilde{Y}^{\prime}}
\end{equation}
\begin{equation}
\mathcal{R}^{\prime}=-\tilde{\mathcal{K}}^{\prime}\cdot\frac{L^{\prime}\tilde{\zeta}+G^{\prime}\tilde{X}^{\prime}+M^{\prime}\tilde{\eta}+H^{\prime}\tilde{Y}^{\prime}}{\tilde{\zeta}\tilde{X}^{\prime}+\tilde{\eta}\tilde{Y}^{\prime}}+\frac{L^{\prime}\tilde{\nu}+J^{\prime}\tilde{X}^{\prime}+M^{\prime}\tilde{\mu}+I^{\prime}\tilde{Y}^{\prime}}{\tilde{\zeta}\tilde{X}^{\prime}+\tilde{\eta}\tilde{Y}^{\prime}}
\end{equation}

\bibliography{manuscript_hfbscattering_revtex}% Produces the bibliography via BibTeX.

\end{document}